\documentclass[useAMS,usenatbib]{mn2e}
\usepackage{graphicx,graphics,amsmath}
\title[Polarisation measurements from black hole binaries in their thermal state]{ Prospect of polarisation measurements from black hole binaries in their thermal state with a scattering polarimeter}
\author[Chandreyee Maitra and Biswajit Paul]{Chandreyee Maitra$^{1,2}$\thanks{E-mail:
cmaitra@rri.res.in;} and Biswajit 
Paul$^{1}$\thanks{E-mail: bpaul@rri.res.in;}\\
$^{1}$Raman Research Institute, Sadashivnagar, Bangalore-560080, India;\\
$^{2}$Joint Astronomy Programme, Indian Institute of Science, Bangalore-560012, India}
\begin{document}

\date{}


\maketitle

\label{firstpage}

\begin{abstract}
X-ray Polarisation measurement is a unique tool which may provide crucial information regarding the emission mechanism and the geometry of various astrophysical sources, like neutron stars, accreting black holes, pulsar wind nebulae, AGNs, Supernova Remnants etc. and can help us to probe matter under extreme magnetic fields and extreme gravitational fields. Although the three other domains of X-ray Astronomy \rmfamily{i.e.} timing, spectral and imaging are well developed, there has been very little progress in X-ray Polarimetry with only one definitive polarisation measurement and a few upper limits available so far. Radiation from accreting black holes in their thermal dominated (High Soft) state is expected to be polarised due to scattering in the plane parallel atmosphere of the disk. Also, special and general relativistic effects in the innermost parts of the disk predicts energy dependent rotation in the plane of polarisation and some distinct signatures which can be used as a probe for measuring the parameters of the black hole like its spin, emissivity profile and the angle of inclination of the system. We present the results from an analysis of expected minimum detectable polarisation from some of the galactic black hole binaries GRO J1655-40, GX339-4, H1743-322, Cyg X-1 and XTE J1817-330 in their thermal dominated state with a proposed Thomson X-ray Polarimeter. A proposal for a scattering polarimeter has been submitted to the Indian Space Research Organization (ISRO) for a dedicated small satellite mission and a laboratory unit has been built. Along with the measurement of the degree of polarisation, the polarisation angle measurement is also important, hence the error in the polarisation angle measurement for a range of detection significance is also obtained.
\end{abstract}


\section{Introduction}
Since the birth of X-ray astronomy, sensitivity/capability of timing, spectral, and spatial observations of the X-ray sources have improved by several orders of magnitude and has provided us with a wealth of information regarding the X-ray sky. X-ray polarimetry however, continues to remain a fairly unexplored domain with only one definitive polarisation measurement available made with OSO 8 in $1976$ \citep {weisskopf1976} of the Crab nebula ($ \mathrm{P} = 19 \pm $ 1.0 per cent) thus proving the synchrotron nature of the radiation. It also provided upper  limits of - $ 13.5 $ and $ 60 $ per cent for the pulse phase averaged degree of polarisation in two X-ray accretion powered pulsars Cen X-3 \& Her X-1 respectively \citep {silver1979}. Although Radio and Optical astronomers extensively use polarimetry to probe into the radiation mechanism and geometry of sources, X-ray astronomy is still behind them in this regard which would provide some crucial information regarding the radiation mechanism, source geometry, inclination and magnetic fields \rmfamily{etc}. Sensitive X-ray polarimetry could also be used as a probe for fundamental physics since it gives information on matter under extreme gravitation or magnetic fields. It could also solve degeneracy between different theoretical models which cannot be solved by timing and spectroscopy studies alone. Various emission mechanisms like synchrotron, non-thermal bremsstrahlung etc. give rise to high degree of polarisation. Other emission processes for polarisation could be scattering of initially unpolarised radiation in asymmetric plasmas or disks. Vacuum birefringence through extreme magnetic fields or the phenomenon of compton dragging can also give rise to polarisation. The potential sources of X-ray polarisation are accretion and rotation powered pulsars, thermal or reflected emission from the accretion disk. Cosmic acceleration sites such as supernovae remnants and jets in Active Galactic Nuclei or microquasars are also sources of polarisation. \newline In recent times there has been significant development in X-ray polarisation measurement techniques \citep {costa2001, bellazini2006, costa2008}. Various X-polarimetry missions are under development. Out of them the most significant one is the approved \emph{GEMS} mission \citep {swank2008} which is based on photoelectron polarimeter technique. Recently a proposal has been submitted to the Indian Space Research Organization (ISRO) for a dedicated small satellite experiment for measuring X-ray polarisation in the $ 5  - 30 $ \rmfamily{kev} range. This will be based on the principle of polarisation measurement by anisotropic T2homson scattering of X-rays and will be sensitive to the bright X-ray sources. A laboratory unit has been built for the same \citep {rishin2009}. Here we calculate the Minimum Detectable Polarisation (MDP) that we would get with the X-ray polarimeter experiment for the thermal emission from the accretion disk for some of the galactic black hole sources during their thermal dominated spectral states. Apart from the MDP, the error in measurement of the angle of polarisation for particular detection significance limits is also calculated. This analysis is of importance due to the following reasons:
\begin{enumerate}
\item Sources with strong X-ray polarisation like accretion powered pulsars or reflection dominated emission from the accretion disk         \rmfamily{etc.}, are expected to have a fairly high degree of polarisation and are sources of hard X-rays which are most suited for observation by the Thomson X-ray polarimeter. Galactic black hole binaries however, have a soft spectra in their thermally dominated state and typically peak at $1 - 3 $ \rmfamily{kev}. There is also expected to be very little flux above $\sim 10$ \rmfamily{kev}. The radiation itself is also not polarised to a high degree except in the innermost parts of the accretion disk in some cases. This makes its detection by the Thomson polarimeter difficult.
\item Black hole binaries have different spectral states and display state transitions in which either the thermal or the non thermal component dominates the X-ray flux. They are also mostly transient sources which go into outburst once in a while during which they make their transition to the thermally dominated soft state. Therefore, detection of polarisation from the black hole binaries in their thermal state would also be constrained by the condition of one of the sources going into outburst/state transition during the lifetime of the mission. 
\end{enumerate}
\section[]{Proposed Experiment}
\subsection{Objective}
A Thomson scattering X-ray polarimeter has been developed and a small satellite mission with a similar payload is under consideration by ISRO. It is based on the principle of anisotropic Thomson scattering of X-ray photons working in the $5-30 $ \rmfamily{kev} (with Lithium scatterer) and $8-30 $ \rmfamily{kev} (with Beryllium scatterer). It has a sensitivity of $2-3$ per cent MDP in a $50-100$ mCrab source for an exposure of one million seconds \citep {rishin2009}. The proposed experiment will be useful in measuring the degree and direction of X-ray polarisation of a few ($ \simeq 50 $) bright cosmic X-ray sources including accretion powered binary X-ray pulsars, galactic black hole candidates, rotation powered pulsars and magnetars, SNRs and Active Galactic Nuclei.
\subsection{Design and Polarisation measurement technique}
Though the most sensitive polarisation measurement devices are based on photoelectron track imaging, they require to be coupled with high throughput X-ray mirrors due to a small detector size. They also cover a softer energy range. The proposed scattering experiment is based on the well established technique of X-ray polarisation measurement using Thomson scattering which has moderate sensitivity over a relatively large bandwidth suited in the energy band of our interest. The experiment configuration consists of a central low Z (Lithium, Lithium Hydride or Beryllium) scatterer surrounded by xenon filled X-ray proportional counters as X-ray detectors which collects the scattered X-ray photons. The instrument is rotated along the viewing axis leading to the measurement of the the azimuthal distribution of the scattered X-ray photons which gives information on polarisation. The sensitivity of this experiment is dependent on a) collecting area b) scattering and detection efficiency c) detector background and d) modulation factor of the instrument.
\subsubsection{Sensitivity and Minimum Detectable Polarisation of the Thomson scattering X-ray Polarimeter}
The experiment is based on the principle that the differential Thomson scattering cross section for polarised radiation has an azimuthal dependence. The expression for differential cross section is given by 
\begin{equation}
 \mathrm{d}\sigma=\mathrm{r}_e^2(1-\mathrm{\sin}^2\theta \ \mathrm{cos}^2\phi)\mathrm{d}\theta \mathrm{d}\phi
\end{equation}
where $ \theta $ is the angle between the direction of the incident and scattered photon, $ \phi $ is the angle between the scattering plane defined by the direction and the electric field vector of the incident photon, $\mathrm{r}_e$ is the classical electron radius. The resultant azimuthal distribution follows the integral of the differential cross section and is given by
\begin{equation}
 \mathrm{C(\phi)} = \mathrm{B} + \mathrm{A\mathrm{\sin}(2\phi - P)}
\end{equation}
A and B are constants and $ \phi $ is the azimuthal coordinate and P the angle of polarisation of the incident radiation.\newline\\The modulation factor $ \mu $ for the pattern is expressed as
\begin{equation}
 \mu  = \frac{\mathrm{C_{max}} - \mathrm{C_{min}}}{\mathrm{C_{max}} + \mathrm{C_{min}}}
\end{equation}
The degree of polarisation for the incident beam is defined as
\begin{equation}
 \mathrm{P_{pol}}= \frac{\mu_p}{\mu_{100}}
\end{equation}
where $ \mu_p $ is the observed modulation factor and $ \mu_{100} $ is the modulation factor of the instrument in the absence of background for a $100$ per cent polarised radiation\newline\\ Minimum Detectable Polarisation (MDP) of the measurement for 99 per cent confidence level is given as \citep {weisskopf2010} -
\begin{equation}
 \mathrm{MDP_{99}}=\frac{4.29}{\mu_{100} \mathrm{S}}\sqrt{\mathrm{\frac{(S + B)}{T}}}
\end{equation}
where S is the total source counts in the units of counts/sec given as $ \mathrm{S}=\mathrm{R}_{src}*\mathrm{A}_c*\epsilon $; $ \mathrm{R}_{src} $ being the rate of incident photons per unit area, $ \epsilon $ is the overall detection efficiency and $ \mathrm{A}_c $ the collecting area. B is the total background counts in the same unit given as $ \mathrm{B}=\mathrm{R}_{bkg}*\mathrm{A}_d $ ; $ \mathrm{R}_{bkg} $ being the total rate of background photons per unit area, $ \mathrm{A}_d $ being the total detector area., T the integration time. (Following M. Weisskopf et al. 2010), For a required detection significance of $\mathrm{n}_{\sigma} = 3$ ($99.73$ percent confidence level) \& $ 5 $ ($99.99994$ per cent confidence level) the MDP is calculated to be -
\begin{equation}
 \mathrm{MDP_{3_{\sigma}}}=\frac{4.86}{\mu_{100} \mathrm{S}}\sqrt{\mathrm{\frac{(S + B)}{T}}}
\end{equation}
\&
\begin{equation}
 \mathrm{MDP_{5_{\sigma}}}=\frac{7.57}{\mu_{100} \mathrm{S}}\sqrt{\mathrm{\frac{(S + B)}{T}}}
\end{equation}
 So for a given observation of particular S and B the parameters T determines the MDP at $ \mathrm{n}\sigma $ level.\newline\\This is the expression for the MDP that we get for one day of observations for T secs integration time. To improve our polarisation detectability we should intergrate it for more number of days . So the net MDP obtained for n days of observation each of T secs for a $5_{\sigma} $ detection significance is given by
\begin{equation}
 \mathrm{MDP_{net}}=\frac{7.57}{\mu_{100} S_{net}}\sqrt{\mathrm{\frac{(S_{net} + B_{net})}{T}}}
\end{equation}
where $ S_{net} $ and $ B_{net} $ are the net source and background counts obtained by integration over n days given by 
\begin{equation} 
 \mathrm{S_{net}}=\sum_{i=1}^n \mathrm{S}_i \quad   \& \quad \mathrm{B_{net}}=\sum_{i=i}^n \mathrm{B}_i 
 \end{equation}
\section{Expected Polarisation from Accreting black holes in their thermal state}
One of the earliest work on the expected polarisation properties of thermal emission from the accretion disk was done by \citep {rees1975}. It predicted linear polarisation with the electric field vector lying in the plane of the disk due to electron scattering in its plane parallel atmosphere. This was predicted according to \citet {chandrasekhar1960} calculations for  plane scattering atmosphere where polarisation degree depends on the angle of inclination of the system, the maximum value being $ \simeq 12 $ per cent for edge on systems. \citet {sunyaev1985} showed that the polarisation of the disk photons depend on the optical thickness of the disk. \citet {connors1980} computed X-ray polarisation features from the standard accretion disk using general relativistic treatment which predicted energy dependent rotation in the plane of polarisation in the inner parts of the accretion disk closer to the black hole. \citet {dovciak2008} extended this work by including the effect of optical depth and scattering atmosphere on the expected polarisation signal. \citet {li2008} showed the use of polarisation information obtained of the thermal emission from the accretion disk to infer the inclination of the inner region of the disk. \citet {schnittman2009} computed X-ray polarisation from accreting black holes in their thermally dominated state using a Monte Carlo ray tracing code in General Relativity. For direct radiation from the disk it reproduces Chandrasekhar's results for the outer part of the disk. There is a decrease in the degree of polarisation and energy dependent rotation in the plane of polarisation in the innermost regions due to general relativistic effects. However, for the returning radiation there would be a rotation in the plane of polarisation by ninety degrees at a transition radius, and an enhancement in polarisation which can provide a good estimate of the black hole spin and also map the temperature of the inner accretion disk and its emissivity profile. 
\section{Analysis and Results}
We have calculated the expected MDP from five galactic black hole candidates during their transition to the soft spectral state in the three energy bands of $6-9$, $9-11$, and $11-15$ \rmfamily{keV}. MDP was computed during the days when the blackbody temperature was highest (hence the source most thermally dominated) during the outbursts/state transitions of the sources. Equation $ 7 $ and $ 8 $ is used to calculate the MDP and $ \mathrm{MDP}_{net} $ in the case of a $ 5 \sigma $ statistical detection. The $ \mathrm{MDP}_{net} $ value is also calculated for a $3 \sigma $ statistical detection given by equation $ 6 \sigma $ only in the $11-15 $ \rmfamily{KeV} range, since it may not be detected for a $ 5 \sigma $ detection in all cases in this band.\newline The values of the parameters used for the calculation are as follows :-\newline $ \mathrm{n}_\sigma =5 $ ; modulation factor for $100$ per cent polarisation for our experiment $ \mu =0.4 $ ; integration time T $ = 43\,200 $ secs ( assuming a duty cycle of $50$ per cent in a low altitude near equatorial orbit). \newline The source counts S and background counts B are obtained  from spectral fitting of the Rossi X-ray Timing Explorer \emph{RXTE}/PCA \citep {zhang2006, jahoda1996} spectra in the $ 3 - 30 $ \rmfamily{KeV} range during the specified period of the outburst/state transition of the source. The Proportional Counter Array (PCA) has five xenon filled proportional counter detectors, sensitive in the range of $2-60$ \rmfamily{KeV}, and has an effective area of $ 6500\; cm^2 $ at $6 $ \rmfamily{KeV}. The \emph{RXTE}/PCA spectral data was fitted using XSPEC to a model consisting of interstellar absorption \citep {morrison1983} plus a multicolor blackbody accretion disk \citep {mitsuda1984, makishima1986} and a power-law component. In addition, an iron line and a Fe absorption edge was applied to some data sets. A systematic error of 0.5 per cent was also added in quadrature with the error. For the source counts $ \mathrm{S} = \mathrm{R}_{src}*\mathrm{A}_c*\epsilon $ , $ \mathrm{R}_{src} $ is obtained from the blackbody count rate obtained by the spectral fitting of the data. $ \mathrm{A}_c = 1017\;  cm^2 $ ( collecting area for our experiment ). Following \citet {cowsik2010}, the energy dependent efficiency $ \epsilon $ is calculated for both Lithium and Beryllium scatterer assuming a disc shaped scattering element with an optical depth of unity for Thomson scattering. The values obtained in the three energy bands of interest for both the materials of the scatterer are tabulated in Table $1$.
\begin{table} 
\caption{Energy dependent efficiency for Li \& Be scatterer }
\label{table:1}
\centering
\begin{tabular}{c  c   c}
\hline \hline
Energy range(Kev) & Efficiency(Li) & Efficiency(Be)\\
\hline 
6-9 & 9.4 per cent & 4.0 per cent\\
9-11 & 16.1 per cent & 8.5 per cent\\
11-15 & 21.7 per cent & 14.4 per cent\\
\hline
\end{tabular}
\end{table}
\newline The background consists of the detector internal background B1. The power-law non thermal component of the flux B2  could also be polarised to a significant fraction having an energy dependence and could provide crucial information on the physics and geometry of the power-law photon emitting regions and also on the properties of the black hole \citep {schnittman2009}. However in this work we have considered this component of the flux as a background, due to the following reasons :
\begin{enumerate}
\item Black hole binaries in their low hard states would give us information on the polarisation of the power-law photon emitting regions \rmfamily{i.e}. the corona and also on the black hole parameters. This is however expected to differ significantly from the polarisation obtained from the thermal photons in the inner parts of the accretion discs obtained in the high soft states. Therefore a difference in the polarisation information obtained in the two states would clearly distinguish between the two components. The power-law photons are likely to have only small energy dependence, and, in particular, a continuous change of the polarisation angle with the energy is typical of the GR effects near the black hole.
\item Moreover in the thermal dominated (high soft) spectral states of the black hole binaries, the emission from the disk dominates over the power-law photon emitting regions and hence the polarisation signal can be thought to come predominantly from the accretion disc of the source. There is however, a possibility of some dilution of the polarisation signal in the thermal states due to the non-thermal power-law component.
\end{enumerate}
 Therefore $ \mathrm{B} = \mathrm{B1} + \mathrm{B2} $. B1 is calculated from the background obtained from the spectral fitting of the PCA spectra with the area scaled for the polarimeter detector . Hence \newline 
\begin{displaymath}
 \mathrm{B2 }= \frac{\mathrm{Counts_{AvgPCA}} {\mathrm{(counts/sec)}}}{\mathrm{A_{PCATotal}}}* \mathrm{A_{xpolTotal}}
\end{displaymath}
\begin{equation}
\mathrm{S_{net}}=\displaystyle\sum_{i=1}^n \mathrm{S_i} \quad  \& \quad \mathrm{B_{net}}=\displaystyle\sum_{i=1}^n \mathrm{\mathrm{B1}_i} + \mathrm{B2}*\mathrm{n} 
\end{equation}
This gives us an estimate of best MDP that we could have achieved by integration during that particular outburst/transition of the source using the X-ray Polarimeter. Calculating the MDP in the three energy bands of interest also enables us to do spectro polarimetry that would be crucial to map the energy dependent rotation in the plane of polarisation in the inner parts of the accretion disk and possibly the transition radius where there would be a ninety degrees rotation in the plane of polarisation. This would be essential to calculate the parameter space of the black holes and give an estimate of its spin and emissivity profile.\newline\\We have also calculated the maximum error on determination of the angle of polarisation for a given $ \mathrm{n}_{\sigma} $ value (statistical detection limit).\\\newline Results for the five black hole candidates for which the MDP is calculated is given below.

\subsection{GRO J1655-40}
\begin{figure}
\centering
\includegraphics[scale=0.4,angle=-90]{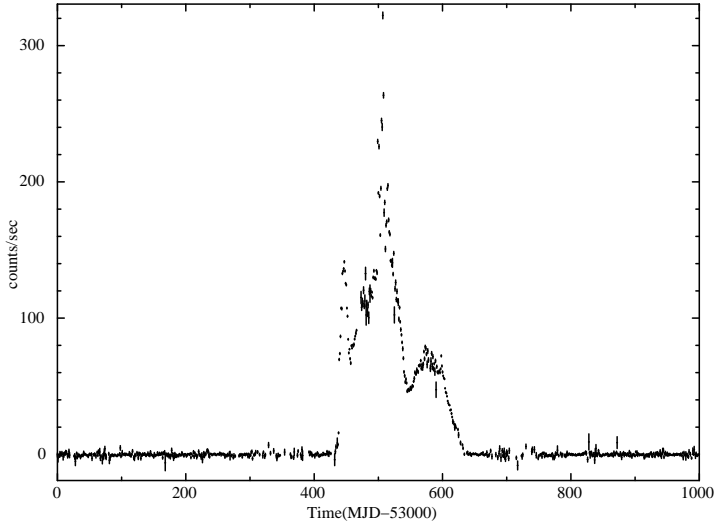}
\caption{ One day averaged long term ASM light curve of GRO J$1655-40 $ showing the $ 2005$ outburst of GRO J$1655-40.$}
\label{fig1}
\end{figure}

\begin{figure}
\includegraphics[scale=0.38,angle=-90]{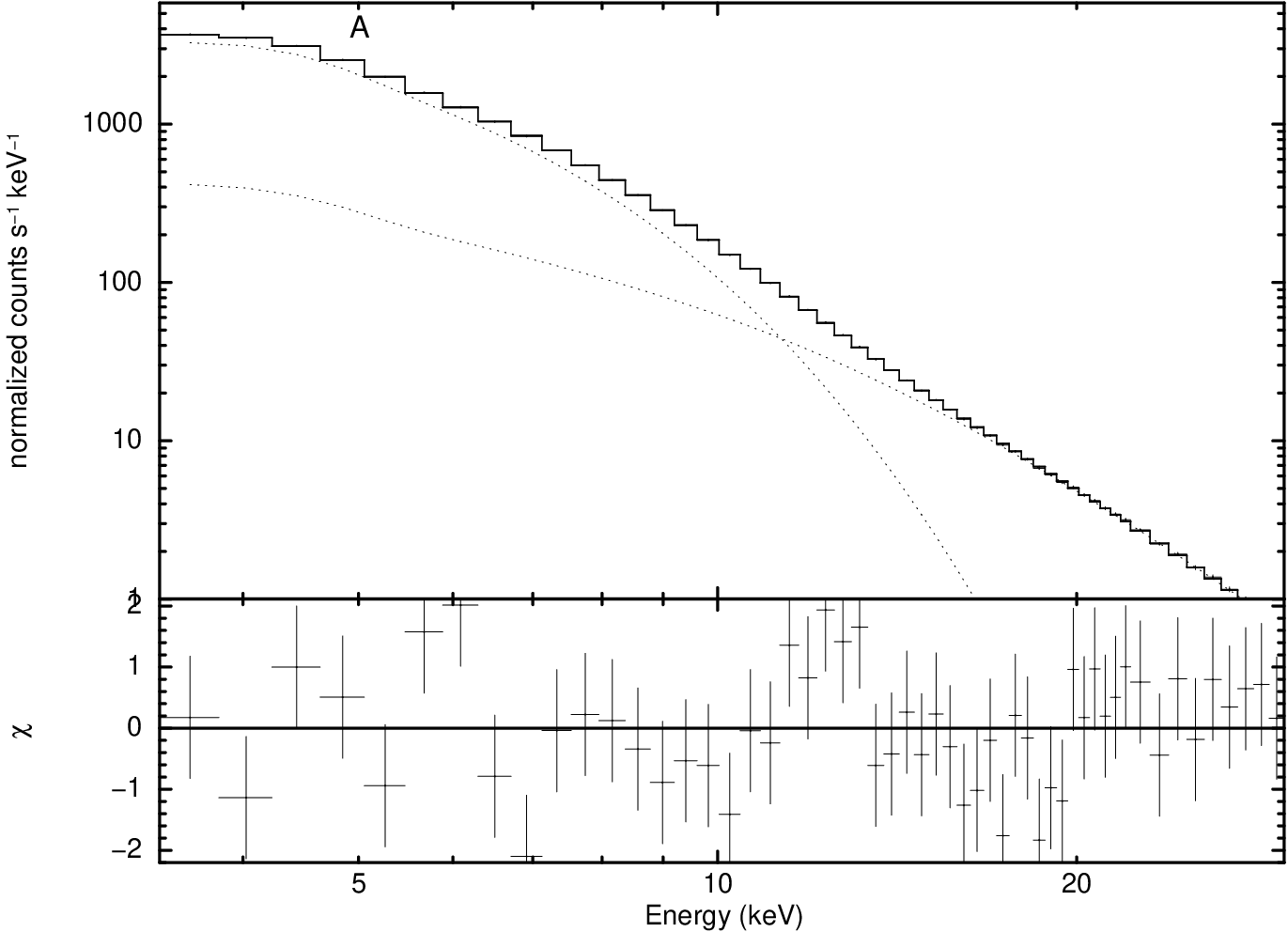}
\includegraphics[scale=0.38,angle=-90]{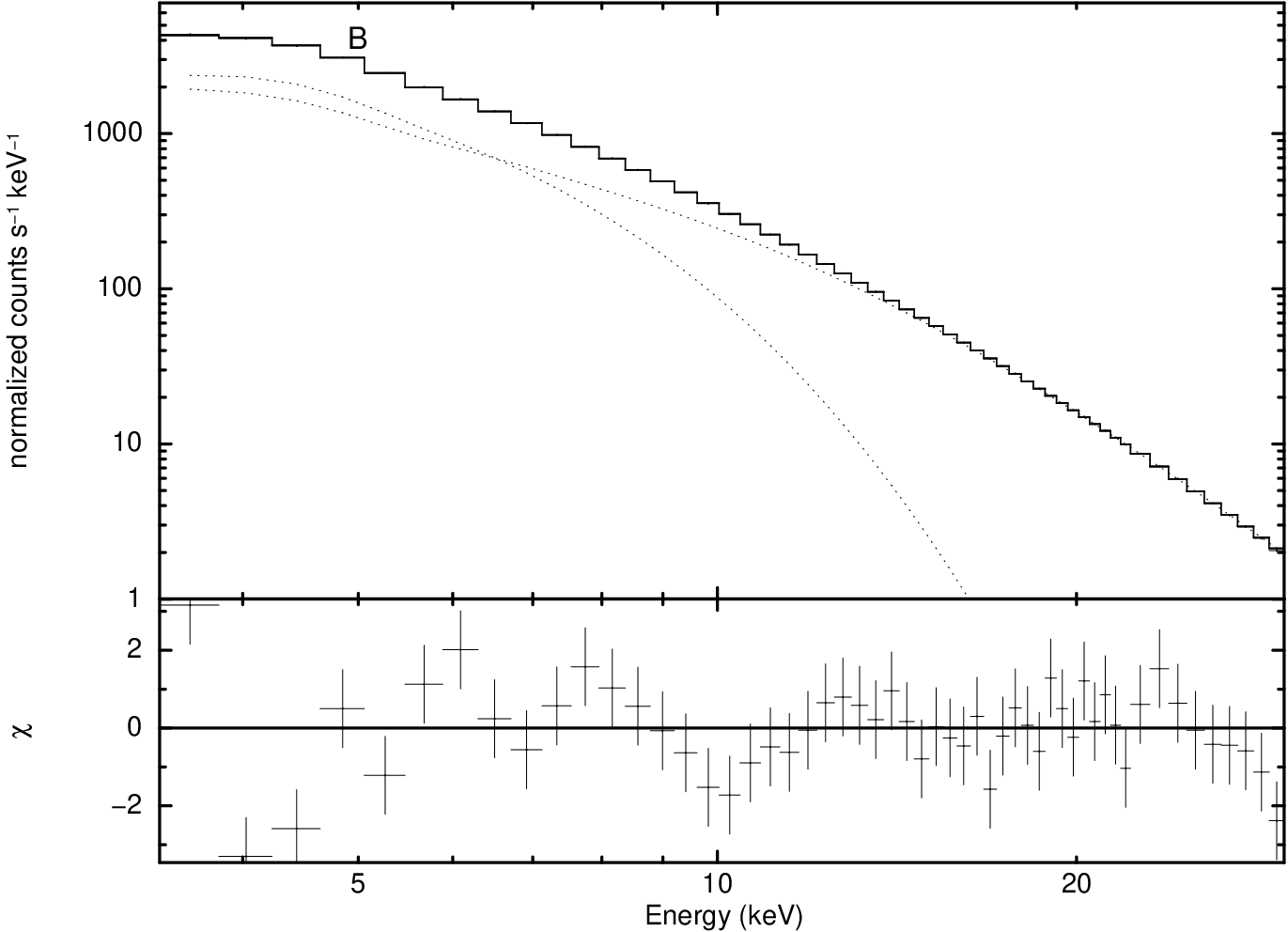}
\caption{Figure A shows the X-ray energy spectrum of GRO J1655-40 in a disc dominated state on MJD $53514$ fitted with the disk blackbody, powerlaw and interstellar absorption. Figure B shows the X-ray energy spectrum of GRO J1655-40 in a power-law dominated state on MJD $53503$ fitted with the disk blackbody, powerlaw  and interstellar absorption}
\label{fig2}
\end{figure}

\begin{figure}
\centering
\includegraphics[width=5cm,angle=-90]{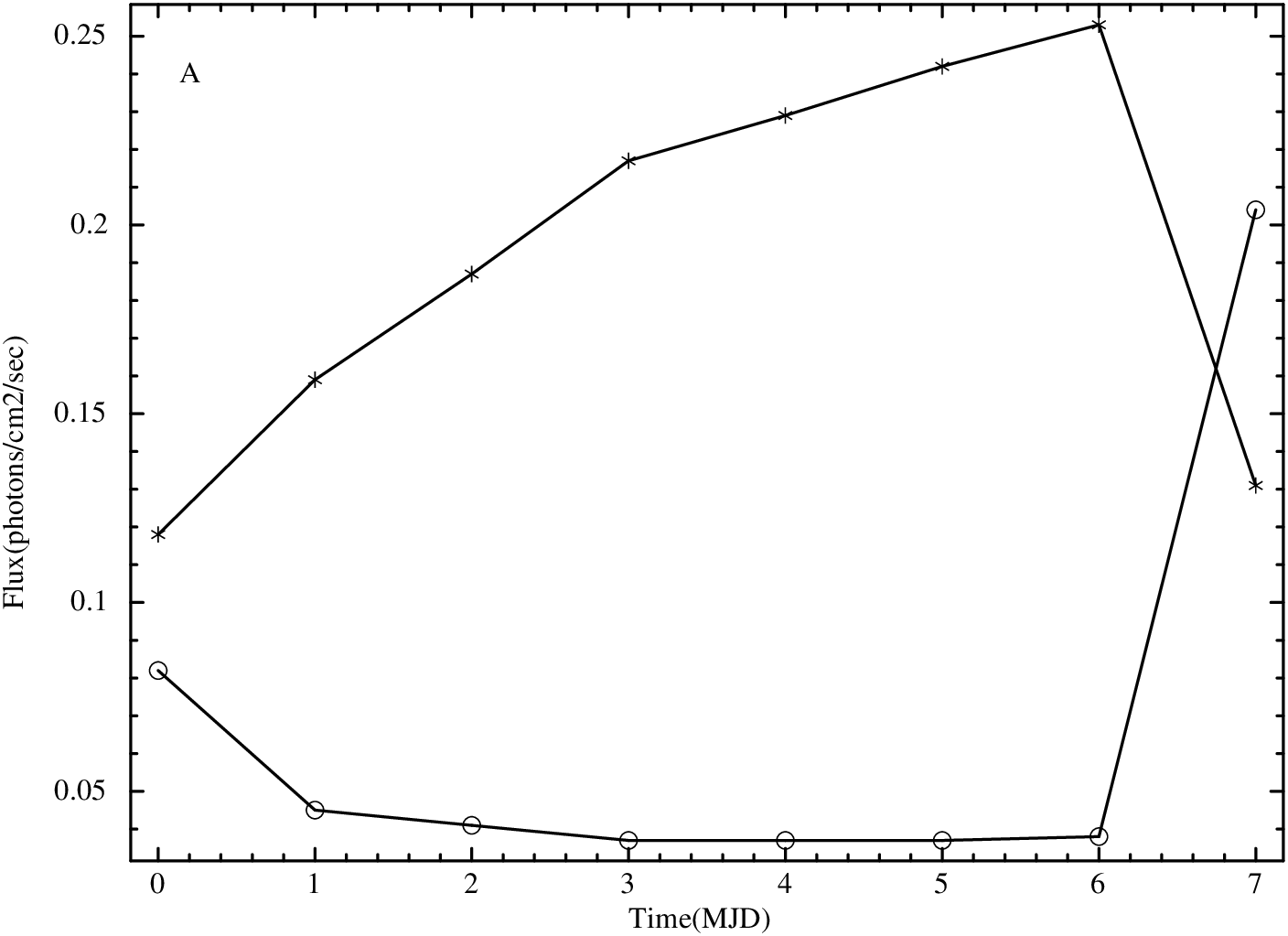}
\includegraphics[width=5cm,angle=-90]{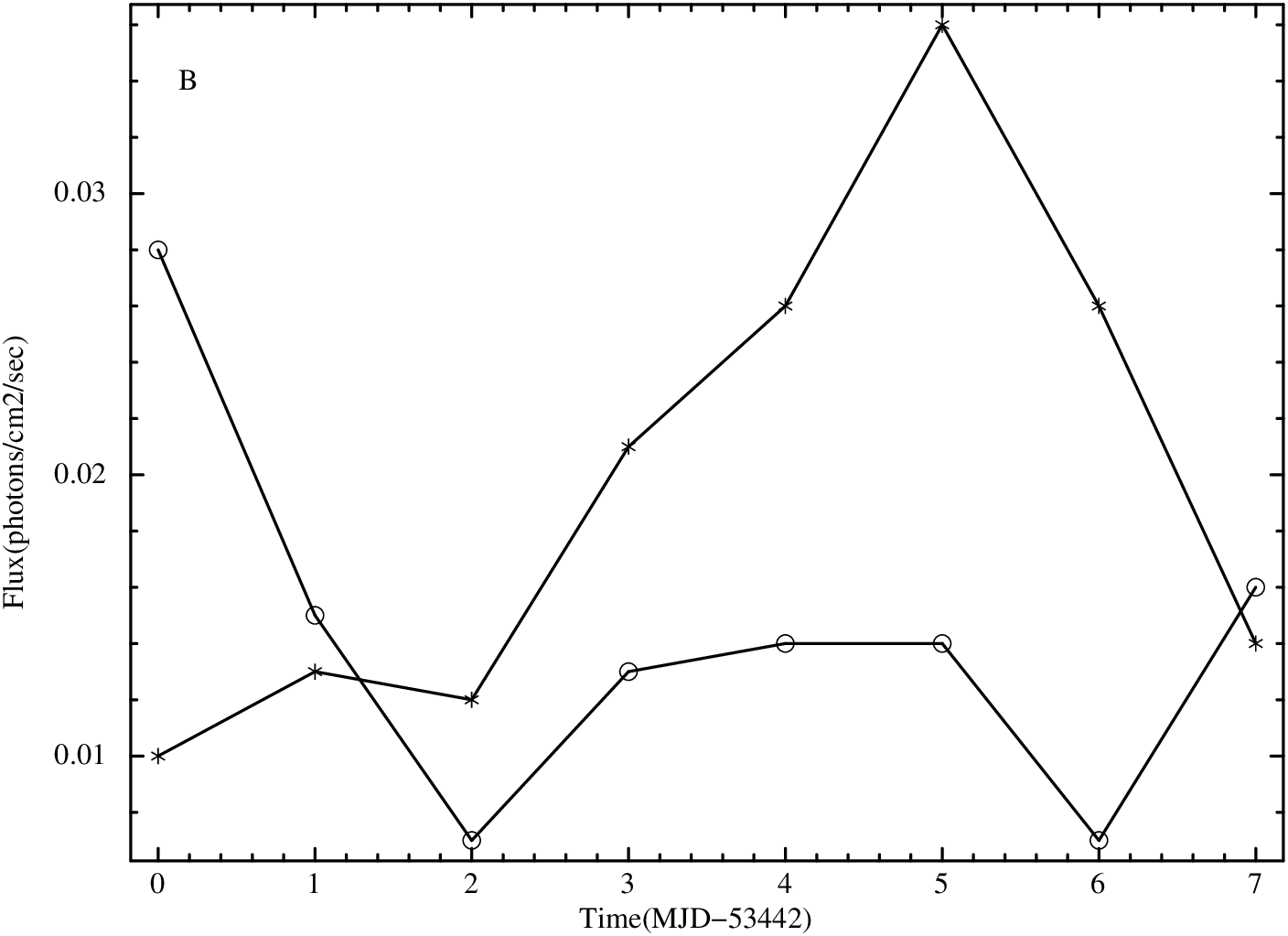}
\includegraphics[width=5cm,angle=-90]{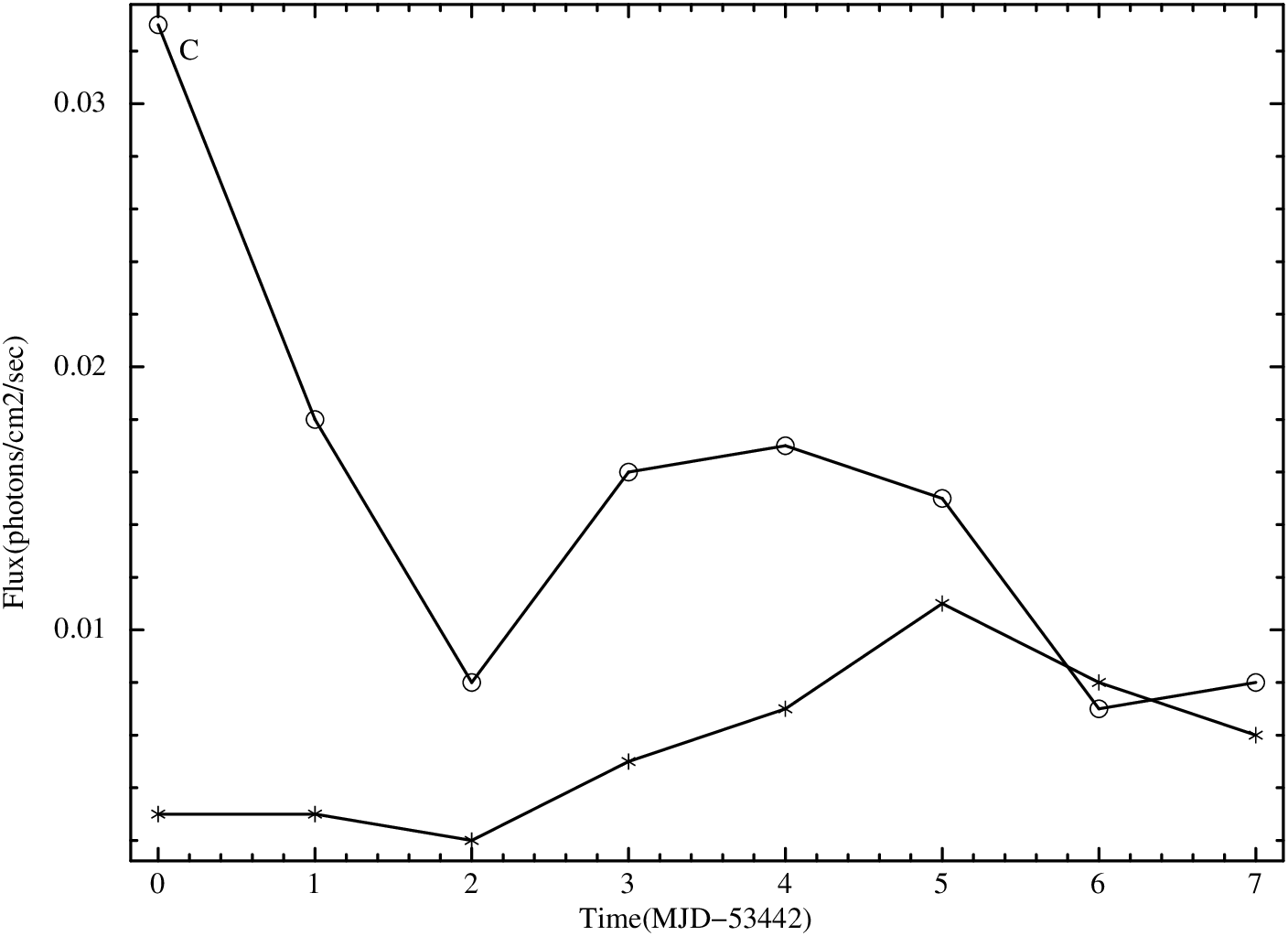}
\includegraphics[width=5.15cm,angle=-90]{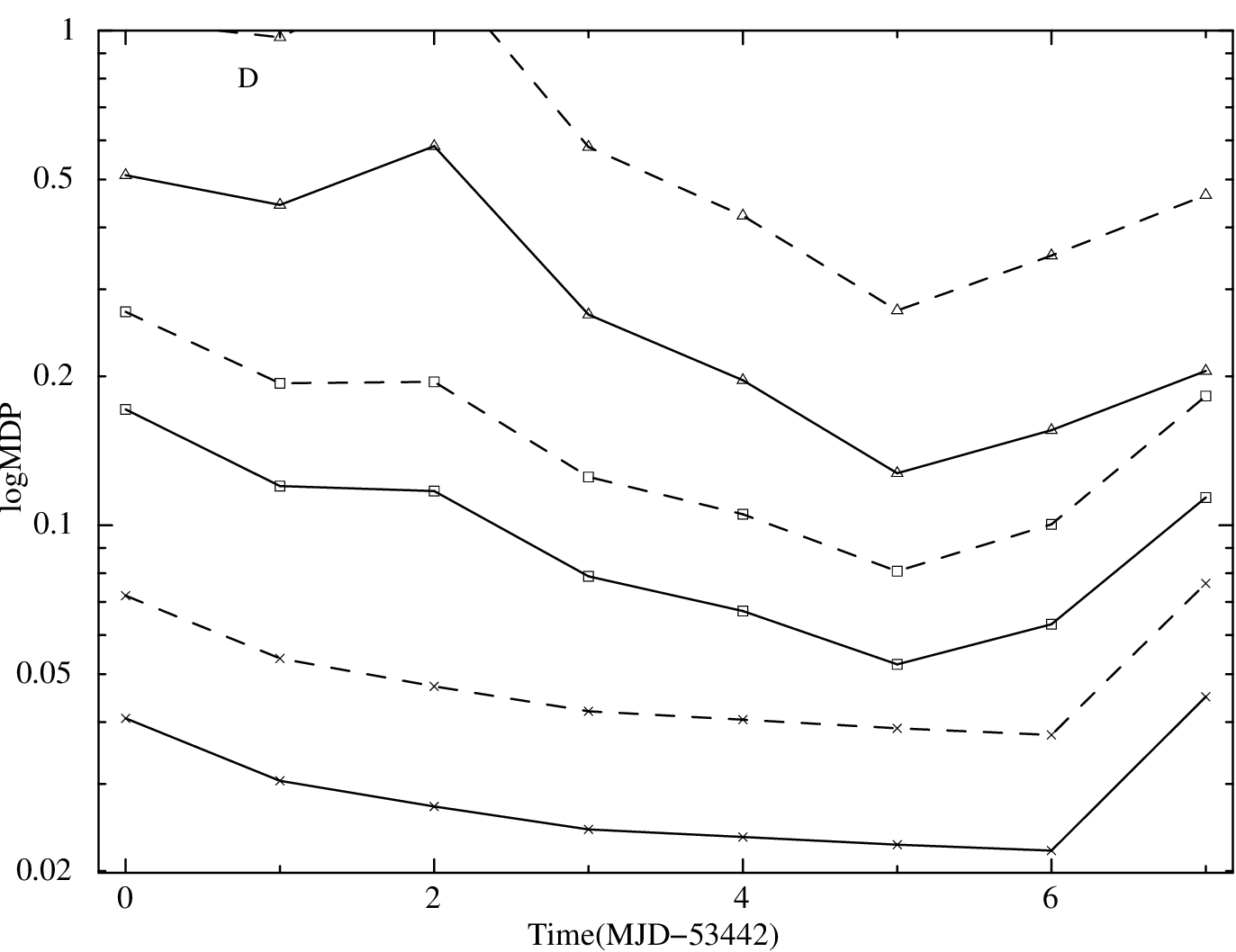}
\caption{Figure A, B, C shows the blackbody and the powerlaw fluxes in the three energy bands for GRO J1655-40 in one stretch of its outburst in $6-9$ \rmfamily{KeV}, $9-11$ \rmfamily{KeV}, \& $11-15$ \rmfamily{KeV} respectively. star indicates the blackbody flux and circle the powerlaw flux. Figure D shows the value of the MDP in the three energy bands for this source during the same time. Cross indicates MDP in $6-9 $ \rmfamily{KeV} square indicates MDP in $9-11$ \rmfamily{KeV} \& triangle indicates MDP in $11-15$ \rmfamily{KeV}.The solid line indicates calculation done with Lithium as a scatterer and dashed line indicates Beryllium as a scatterer.}
\label{fig3}
\end{figure}

\begin{figure}
\centering
\includegraphics[width=5.5cm,angle=-90]{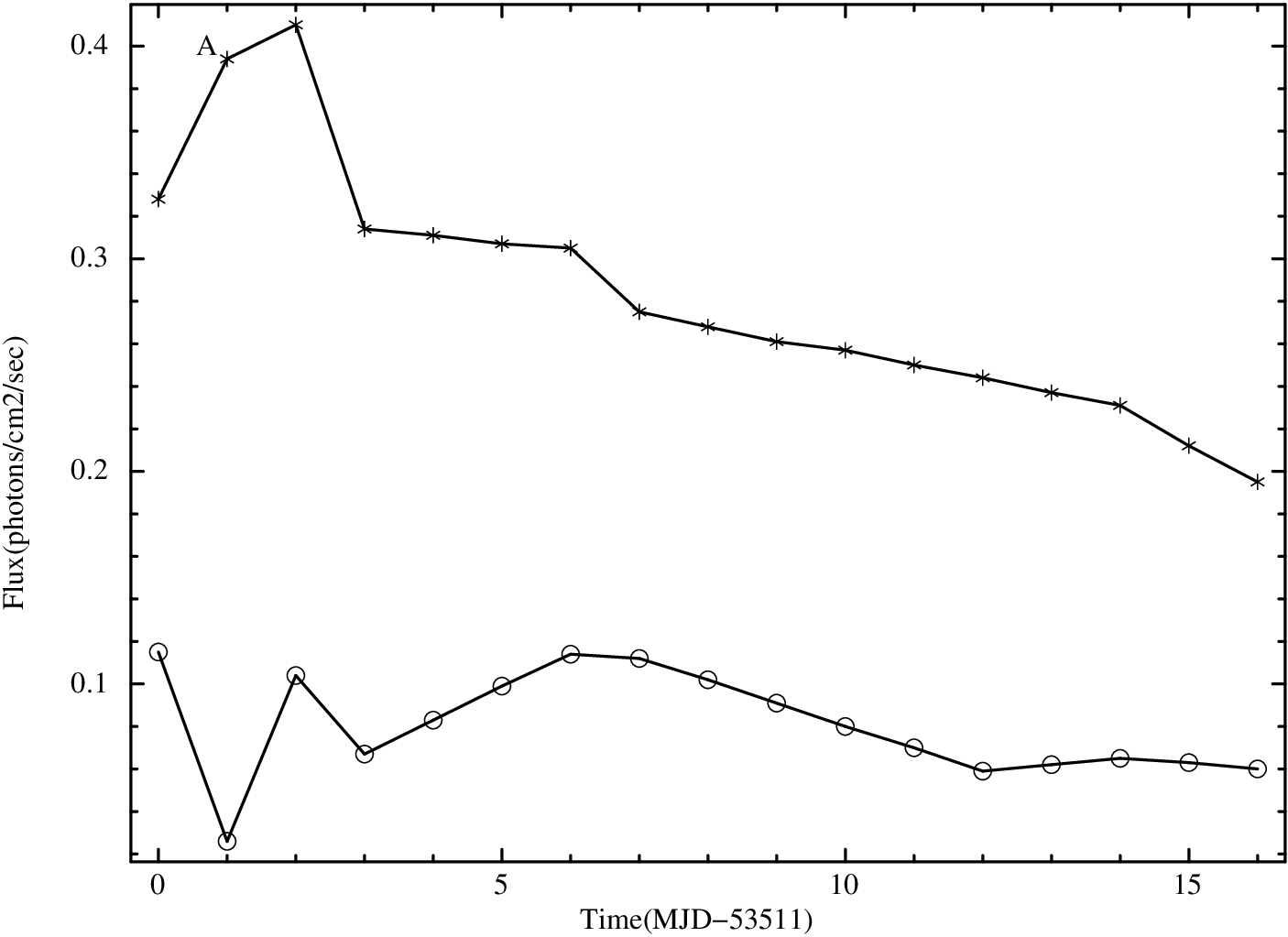}
\includegraphics[width=5.5cm,angle=-90]{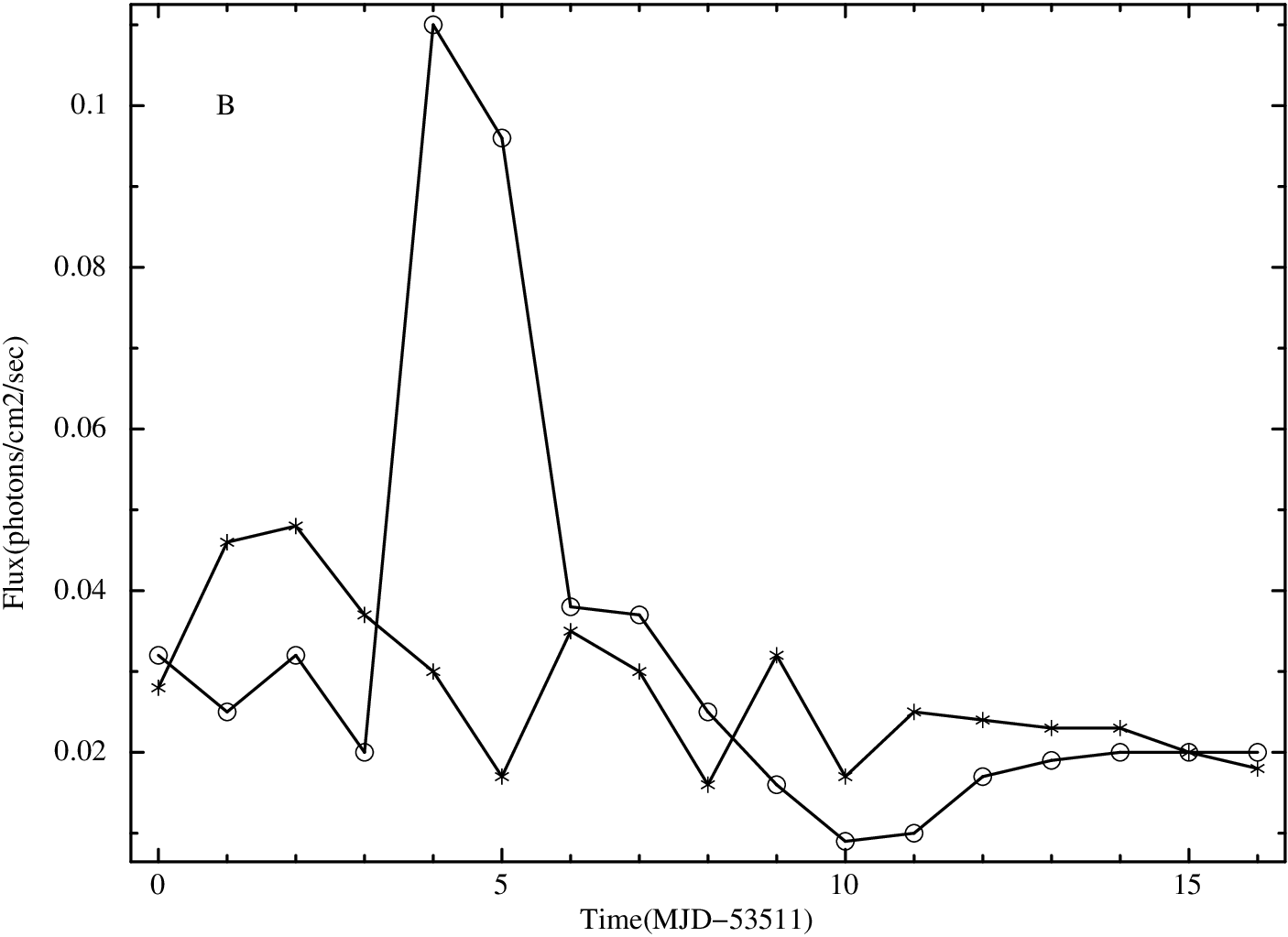}
\includegraphics[width=5.5cm,angle=-90]{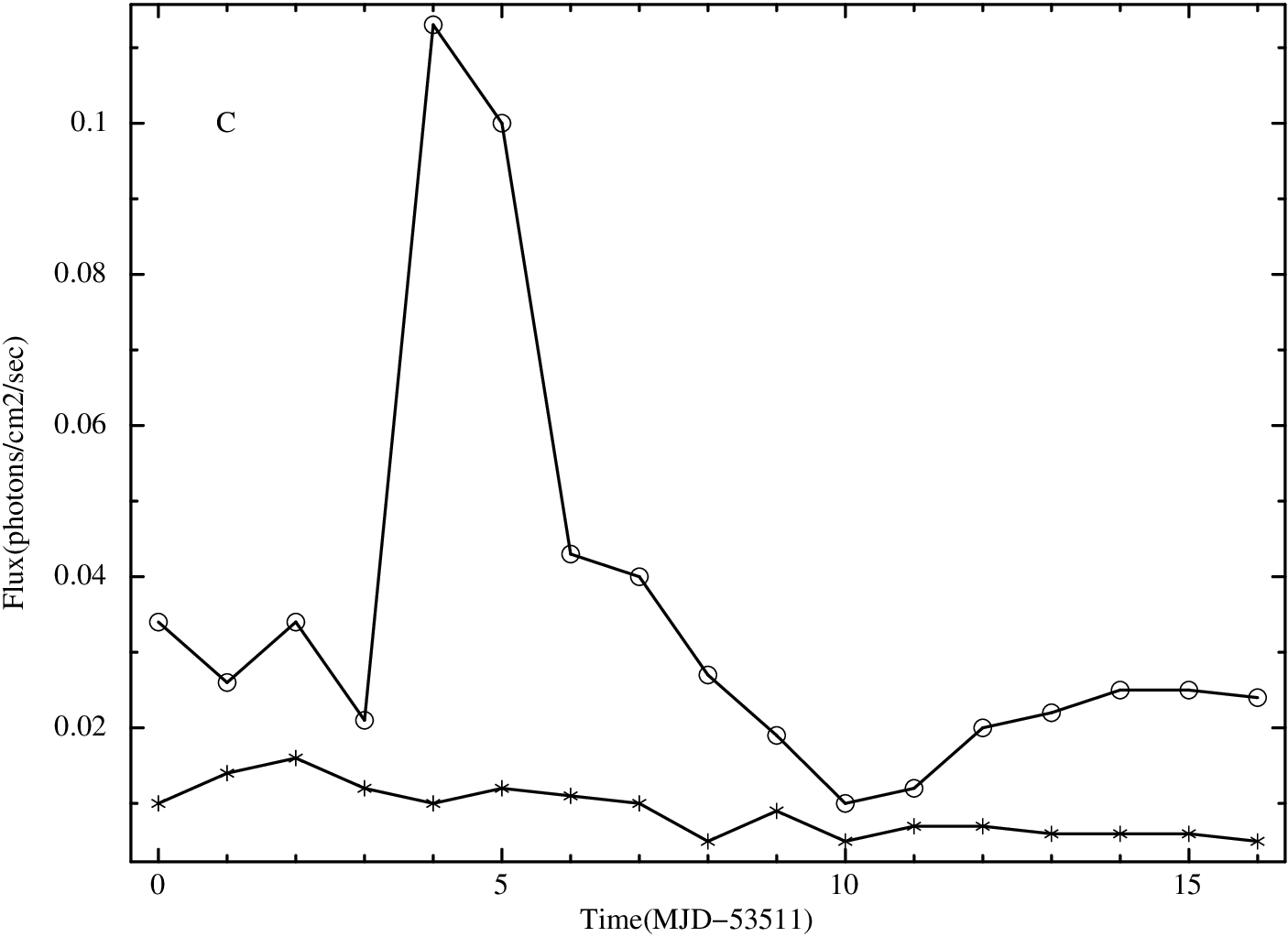}
\includegraphics[width=5.5cm,angle=-90]{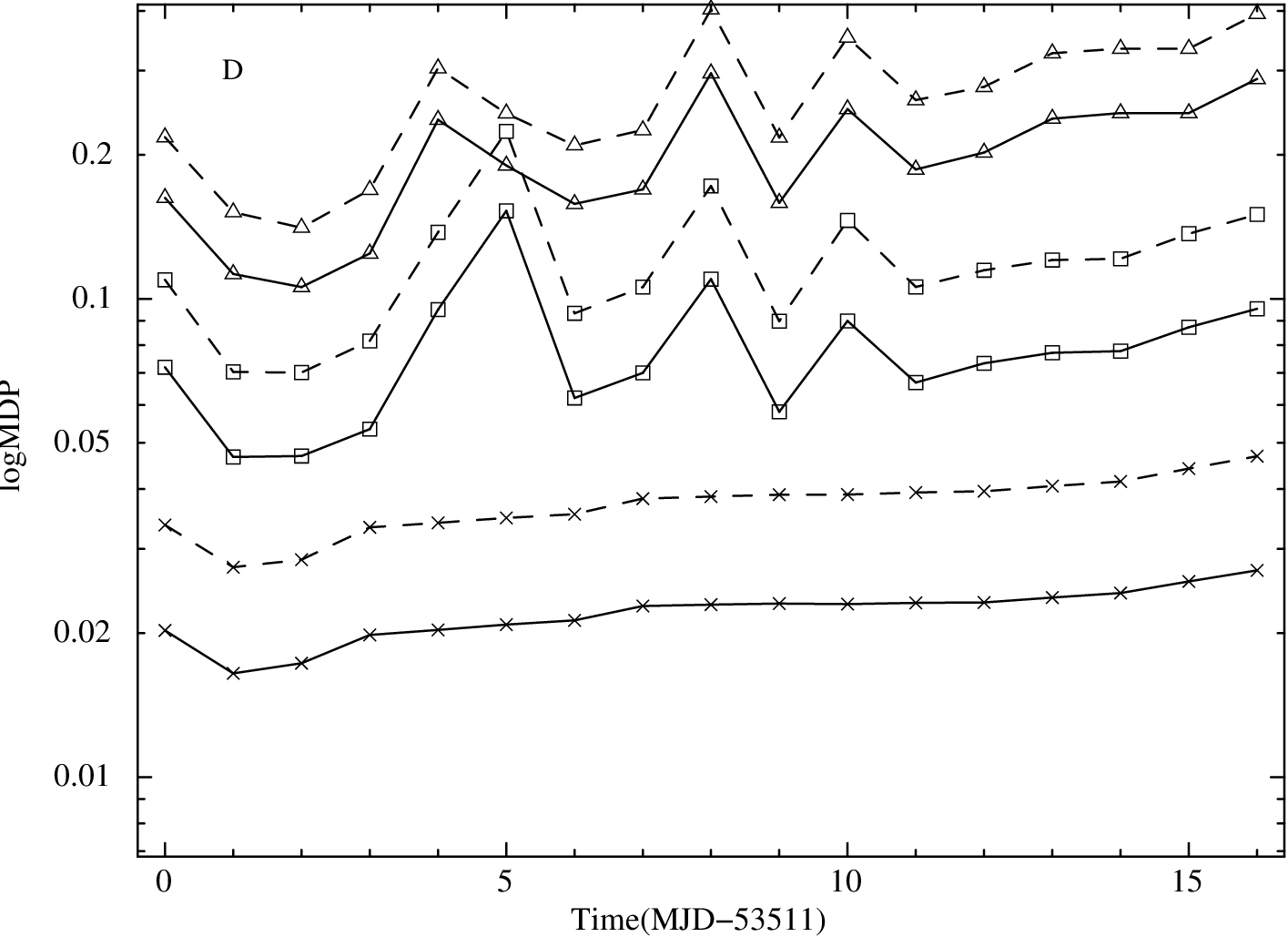}
\caption{ Same as Figure 3 for the second stretch of its outburst of GRO J1655-40 in the year $2005$}
\label{fig4}
\end{figure}

GRO J1655-40 in an X-ray transient which was first discovered in July $1994$ when it went into an outburst and was detected by the Burst and Transient Source Experiment (BATSE) on board \emph{CGRO} in the $20-100$ \rmfamily{KeV} band. \citep {harmon1995}. It is a Low Mass X-ray Binary (LMXB) having the mass of the compact object $ \mathrm{M} = 7.02 \pm 0.22 M_{\odot} $ \citep {orosz1997}. It has also shown signatures of ejection of superluminal jets. It showed an increased X-ray activity in the late February $2005$ as it entered into a new outburst \citep {markwardt2005}. As the outburst evolved the X-ray spectrum passed through various spectral states of low/hard, high/soft, very high states. Figure~\ref{fig1} shows the one day averaged All Sky Monitor ( ASM ) daily light curve of this source during the outburst in the $1.5-12$ \rmfamily{KeV} energy band.\newline\\For the MDP calculation the data was analysed in two stretches MJD $ 53\,442 - 53\,449 $ \&  MJD $ 53\,511-53\,525 $, when the spectrum was found to be the most blackbody dominated in all the three energy bands of our interest. Figure~\ref{fig2} shows the best-fitting spectrum of the source on two days. The first spectrum is of MJD $53\,514$, when the source was blackbody dominated and was hence included in the analysis of the MDP. The second spectrum is of MJD $53\,503$ which is mostly power-law dominated in the energy bands of our interest and hence have not been added in the analysis. Figure~\ref{fig3} \& \ref{fig4} shows the two components (thermal \& non-thermal ) of the flux during the outburst and also the MDP in the three energy bands for the two stretches analysed during the outburst. Although the MDP calculated on individual days for the source may exceed 100 per cent on some days (which means that polarisation in the source cannot be detected with a $5 \sigma $ statistical detection on those days), specially in the $11-15$ \rmfamily{KeV} band and in some days in the $9-11$ \rmfamily{KeV} band, the net MDP obtained by integrating over all the days can be constrained to a much smaller and reasonable value. However in the Figs showing the MDP calculated, the days exceeding 100 per cent MDP have not been plotted. 

\clearpage

\subsection{GX 339-4}
\begin{figure}
\centering
\includegraphics[scale=0.40,angle=-90]{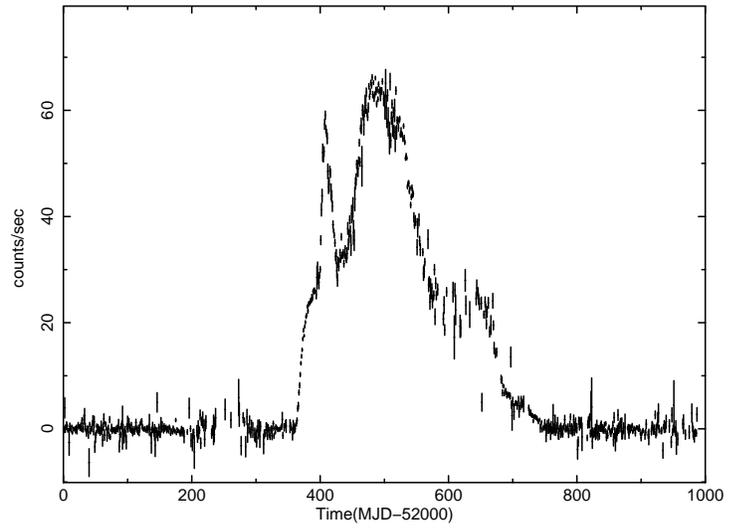}
\caption{ One day averaged long term ASM light curve of GX $339-4 $ showing the $ 2002 $ outburst of GX $339-4.$}
\label{fig5}
\end{figure}

\begin{figure}
\centering
\includegraphics[width=5.5cm,angle=-90]{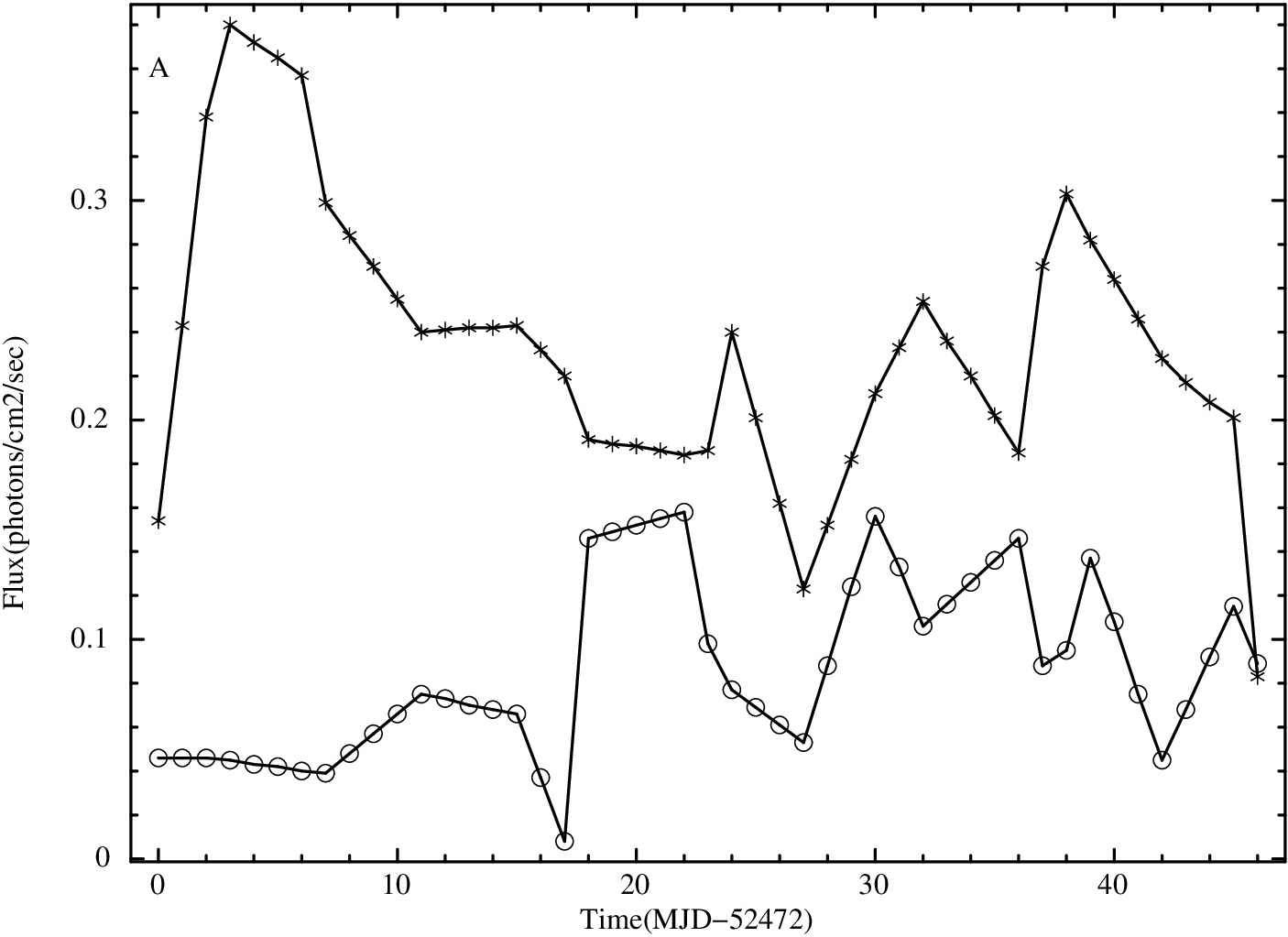}
\includegraphics[width=5.5cm,angle=-90]{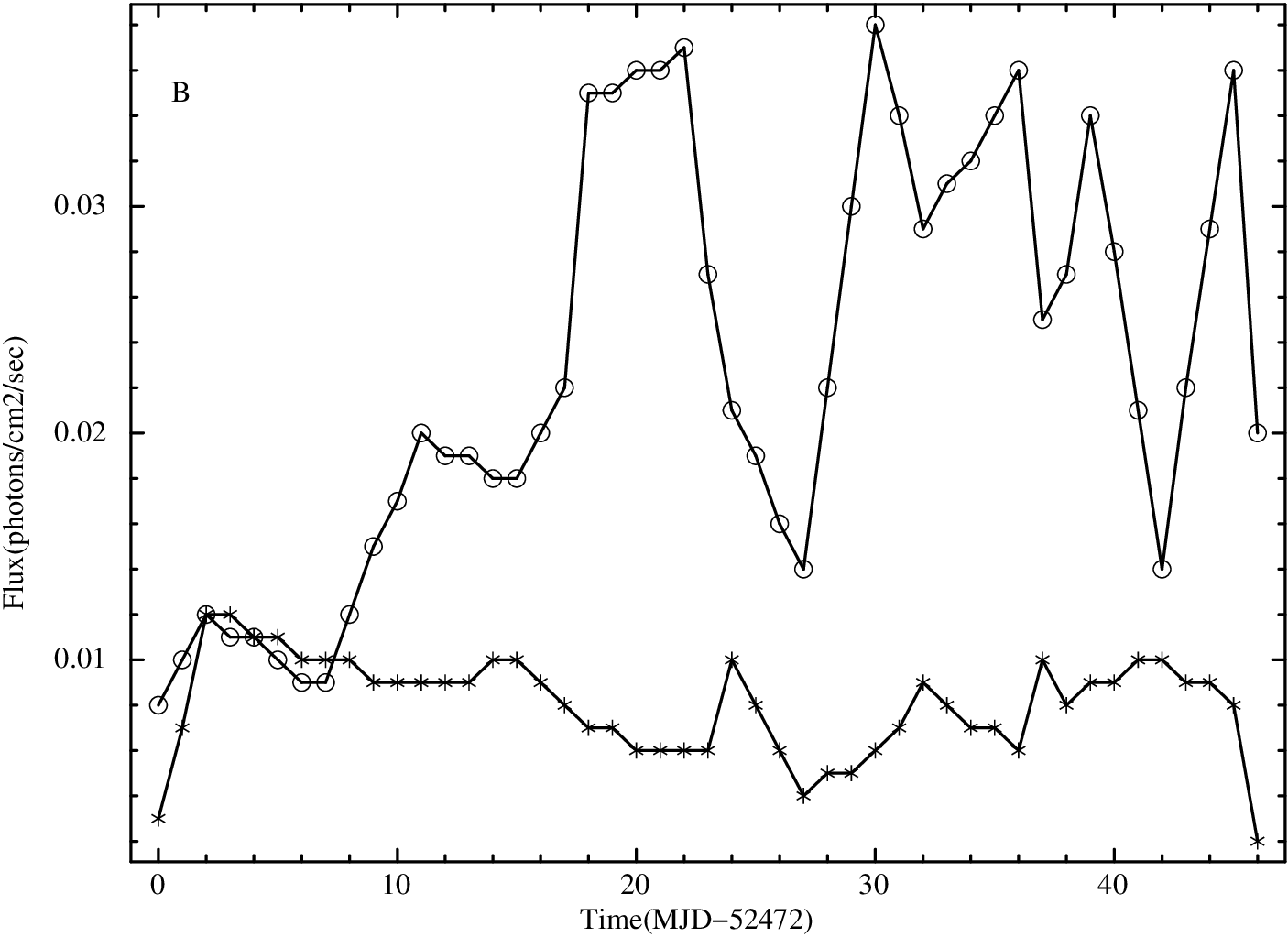}
\includegraphics[width=5.5cm,angle=-90]{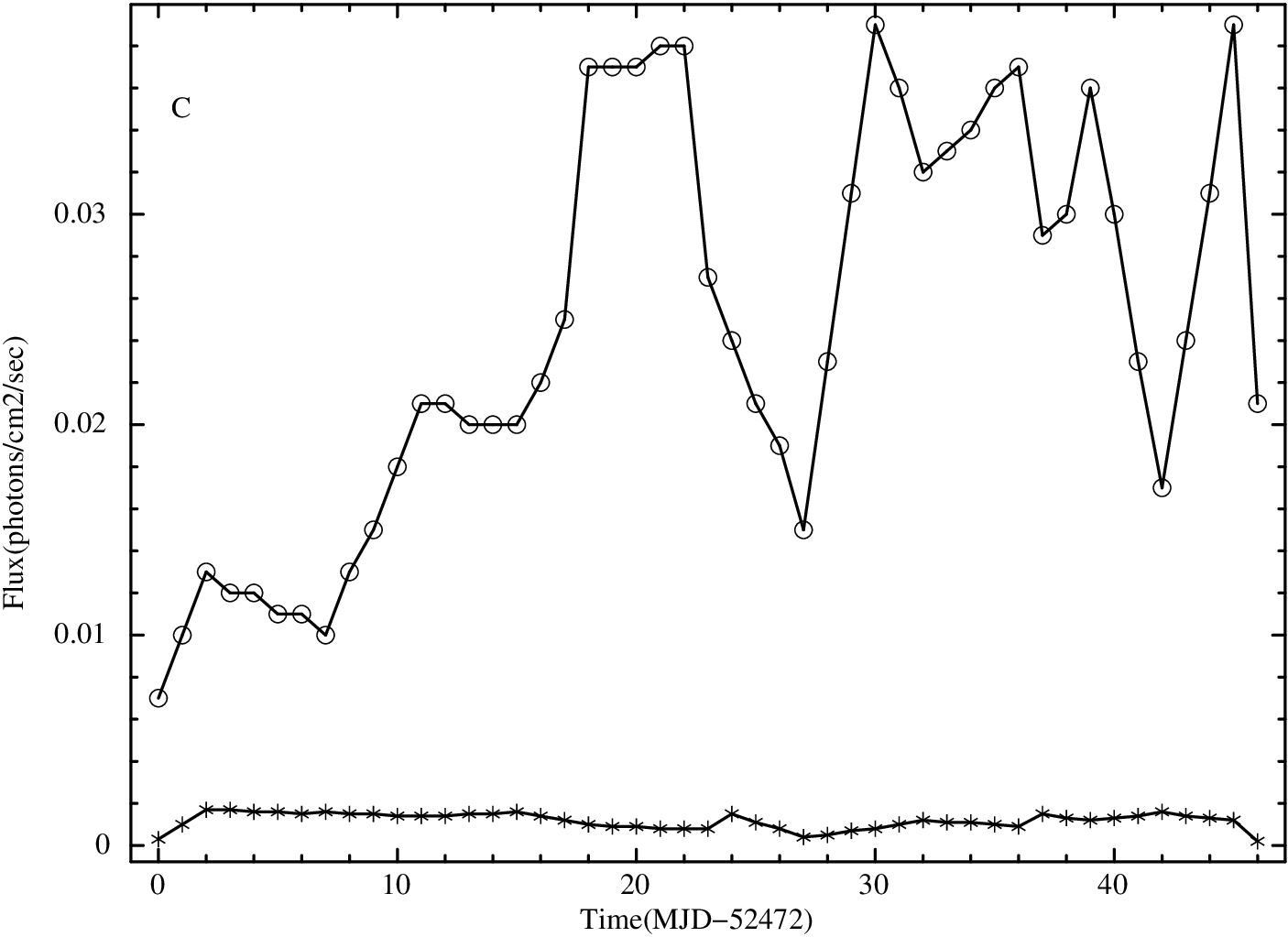}
\includegraphics[width=5.55cm,angle=-90]{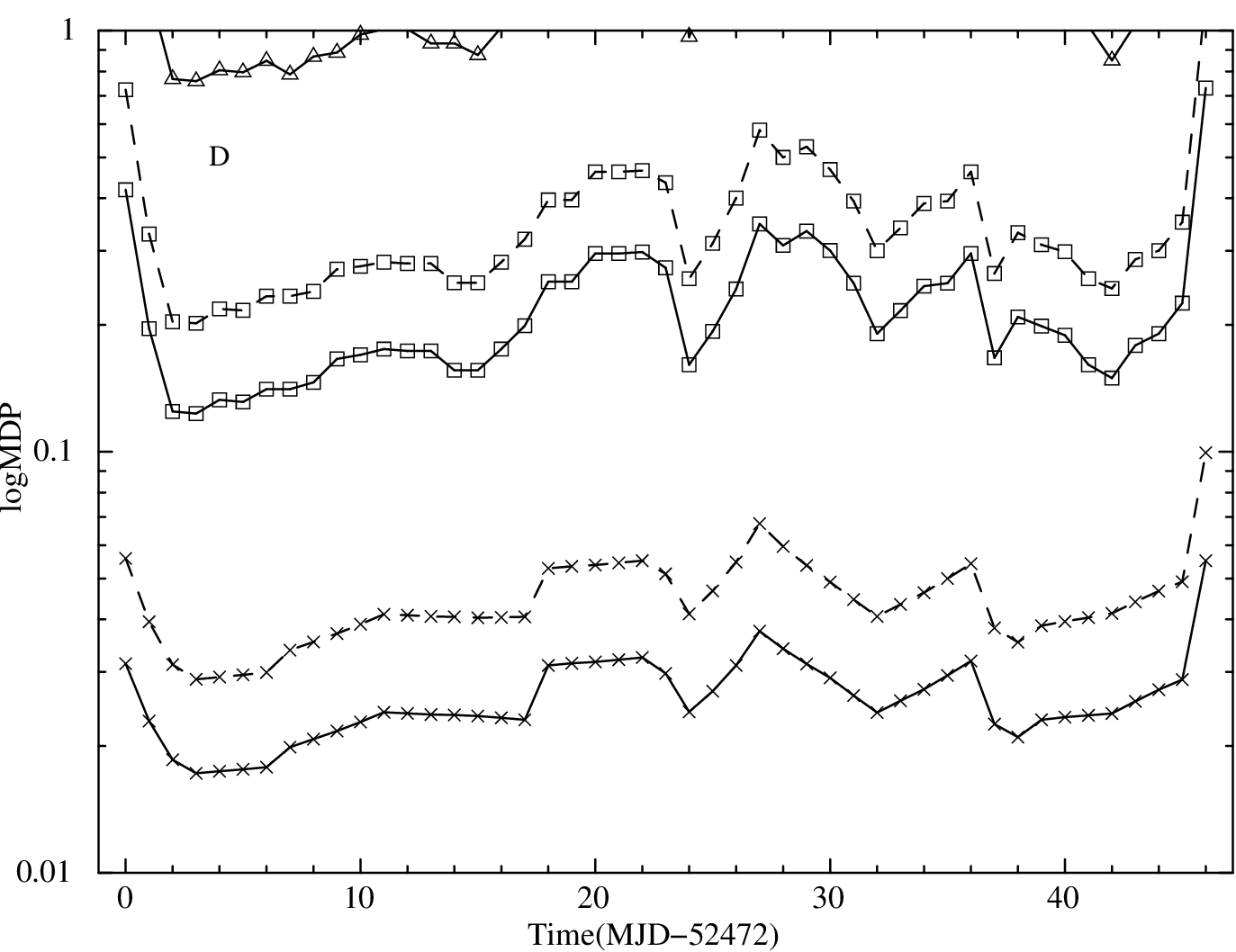}
\caption{ Same as Figure 3, for the source GX339-4 for the outburst in the year $2002$}
\label{fig6}
\end{figure}

GX 339-4 is a recurrent dynamically constrained black hole candidate. It was first discovered with \emph{OSO7} in $1972$ \citep   {markert1973a, markert1973b}. Since then it has undergone several outbursts during which it passed through the various X-ray spectral states. The mass of the compact object is $ \ge 5.8 \mathrm{M}_{\odot} $ \citep {hynes2003}. Radio jets have also been observed in the system. It entered into one of its outbursts which started early in $2002$ April \citep {fender2002, belloni2003, belloni2005}. The X-ray spectrum passed through its various spectral states of low/hard intermediate and then into the high/soft state which it entered beyond MJD 52\,411. The one day averaged daily ASM light curve of this source during the outburst is shown in Figure Figure~\ref{fig5}.\newline We have analysed the data in the stretch of  MJD $ 52\,472-52\,518 $ to calculate the MDP. The net MDP was found to be the lowest in this case $ < 1 $ per cent since apart from this being a very bright outburst, the source was also integrated for a long stretch $ \simeq $ $47$ days. The comparison of the thermal and non thermal flux and the MDP obtained in the three energy bands is shown in Figure~\ref{fig6}. 

\subsection{CYGNUS X-1} 
\begin{figure}
\centering
\includegraphics[scale=0.4,angle=-90]{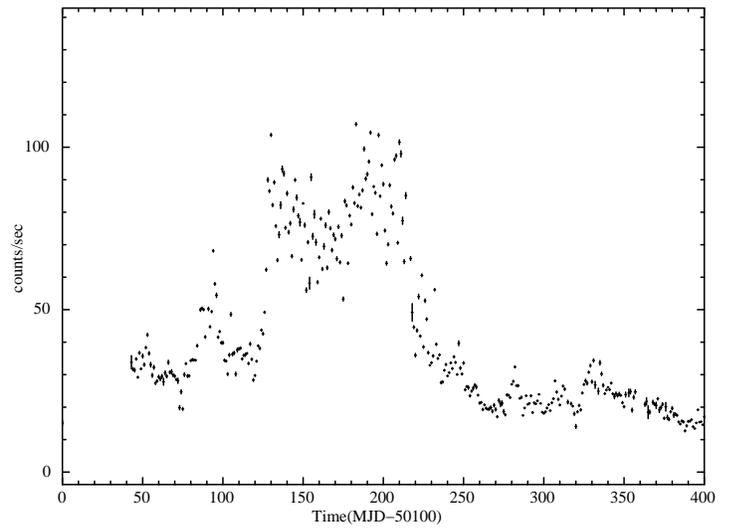}
\caption{ One day averaged long term ASM light curve of Cygnus X-$1$ showing the $ 1996 $ soft state transition of Cygnus X-$1$}
\label {fig7}
\end{figure}

\begin{figure}
\centering
\includegraphics[width=6cm,angle=-90]{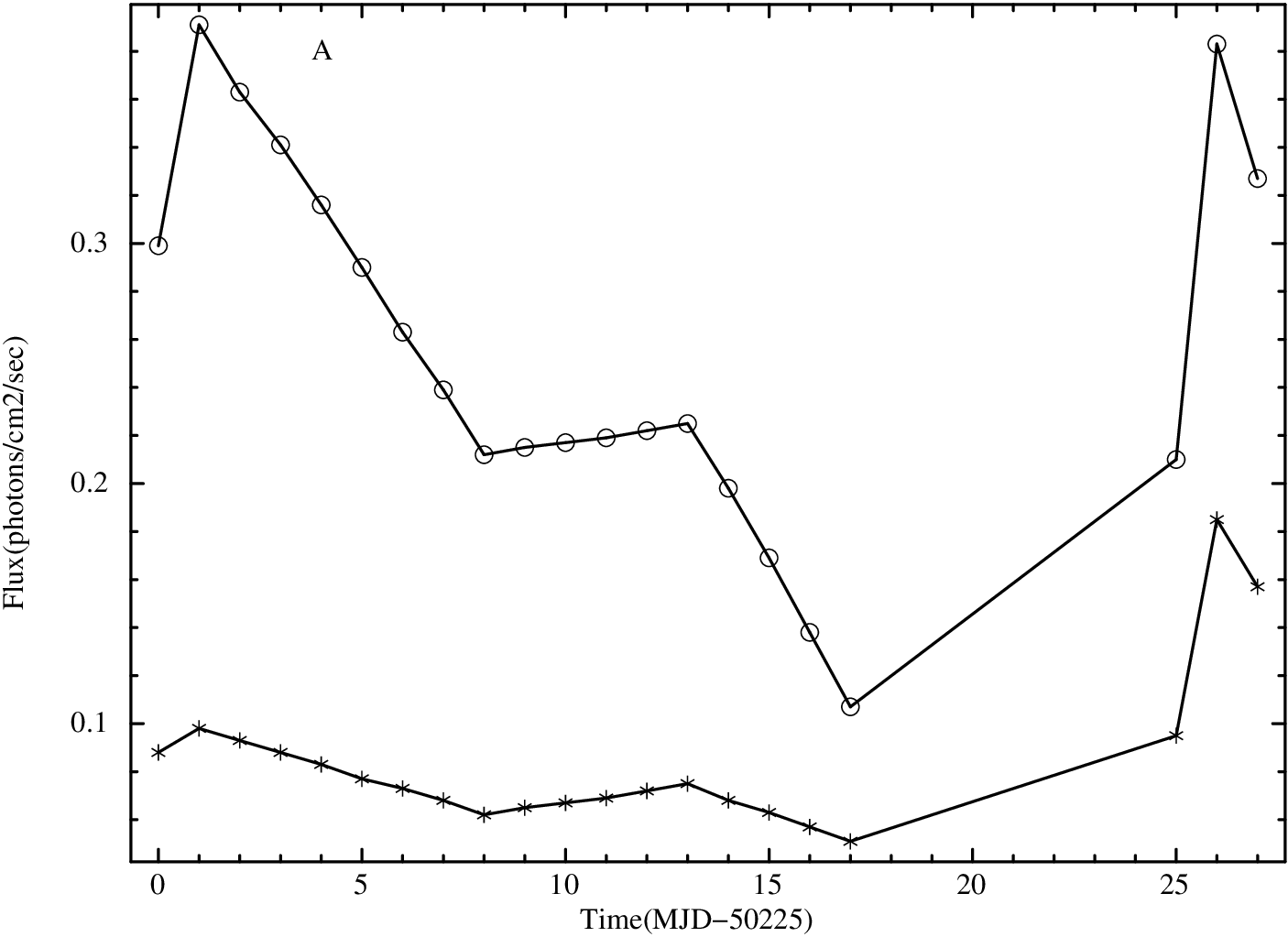}
\includegraphics[width=6cm,angle=-90]{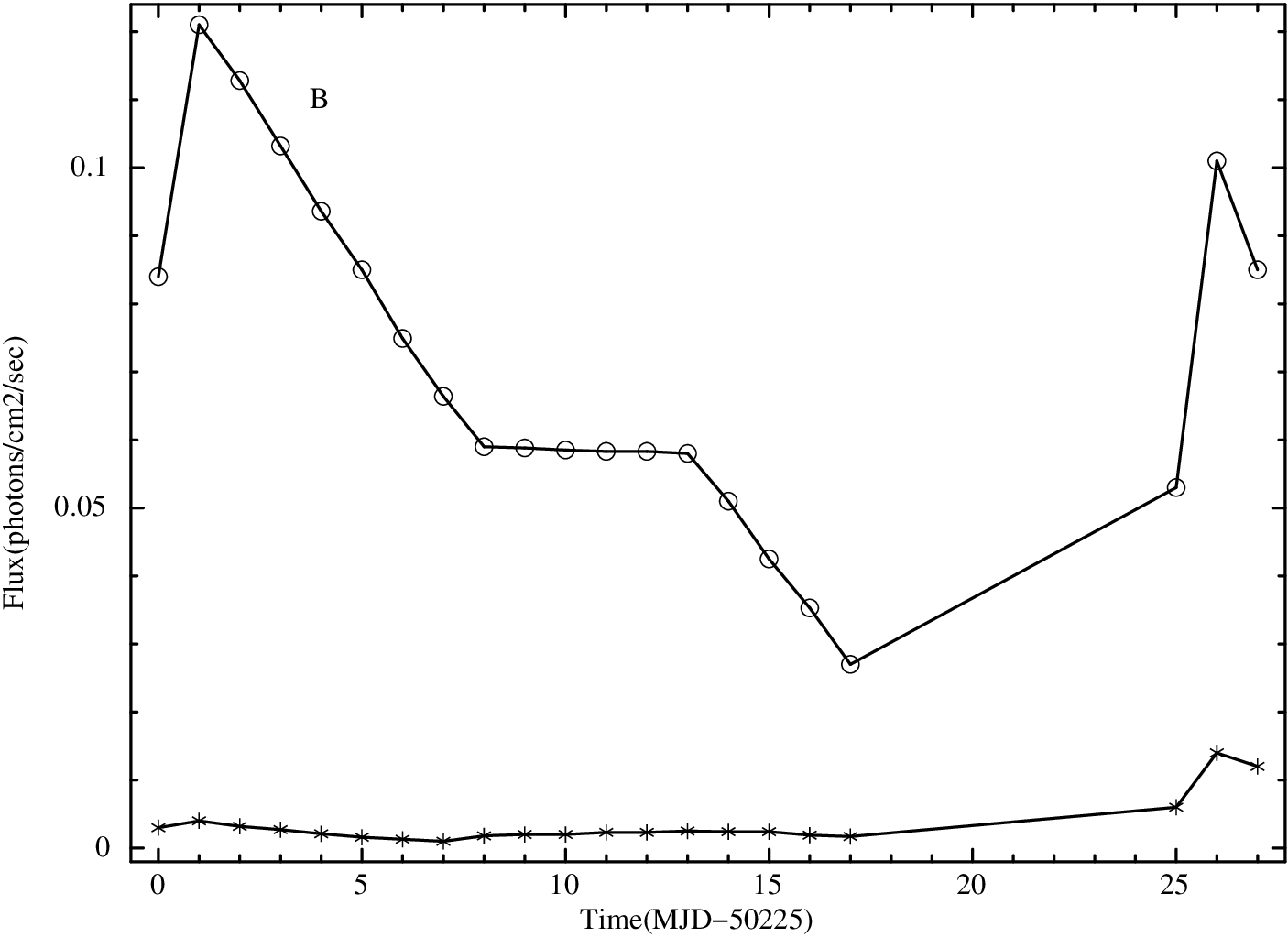} 
\hspace{25cm}
\includegraphics[width=6.15cm,angle=-90]{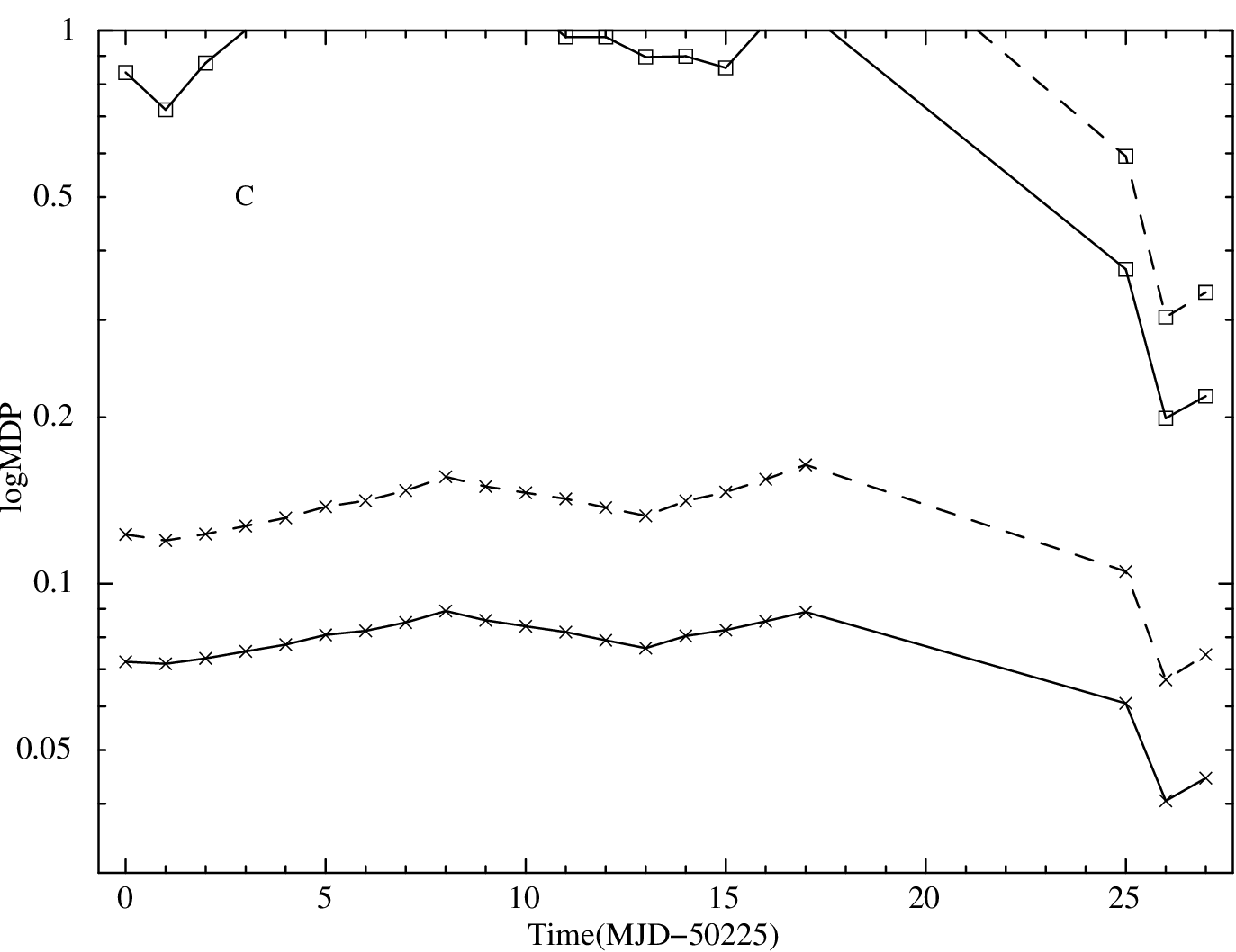}
\caption{Figure A, B shows the blackbody and the powerlaw fluxes in the two energy bands for Cygnus X-1 during its soft state transition in $6-9$ \rmfamily{KeV}, \& $9-11$ \rmfamily{KeV} respectively. Star indicates the blackbody flux and circle the powerlaw flux. Figure C shows the value of the MDP in the two energy bands for this source on each day. Cross indicates MDP in $6-9 $ \rmfamily{KeV} square indicates MDP in $9-11$ \rmfamily{KeV}. The solid line indicates calculation done with Lithium as a scatterer and dashed line indicates Beryllium as a scatterer. The thermal emission from Cygnus X-1 is very weak in the $11-15$ \rmfamily{KeV} band and is not shown here.}
\label {fig8}
\end{figure}

Cygnus X-1 is one of the brightest X-ray sources in the sky and is the most extensively investigated object among all black hole candidates. It was first discovered in a $1964$ June rocket flight \citep {bowyer1965}. The companion mass is in a range between $ \simeq $ 15 $ \mathrm{M}_{\odot} $ \citep {herrero1995} \& $ 30 \mathrm{M}_{\odot} $ \citep {gies1986}. This mass together with the mass function give an estimate of compact object mass between $ \simeq  6.5 $ \& $ \simeq 20 M_{\odot} $ The X-ray spectrum undergoes a transition between the low/hard and the high/soft one. Most of the time the source is in the hard state but it makes occasional transitions to the soft state which is dominated by a soft blackbody like component. During $1996$ it made a complete transition form the low/hard to the high/soft state \citep {cui1996a, cui1996b, zhang1997}. However the blackbody temperature was not very high and the flux increase from its low/hard was only moderate as compared to the high/soft state transitions of the other transient black hole candidates. Figure~\ref{fig7} shows the one day averaged ASM daily light curve of the source during it's soft state transition.\newline\\The data analysed to calculate the MDP ranged between MJD $ 50\,225 - 50\,242 $ \& $ 50\,250 - 50\,252 $. The $ \mathrm{MDP}_{net} $ integrated over all the days is higher than the upper limit on the polarisation obtained by OSO-8 \citep {long1980} from Cygnus X-1 except in the $6-9$ \rmfamily{KeV} with a Lithium scatterer. Therefore polarisation of the thermal component of Cygnus X-1 most probably cannot be observed with the Thomson polarimeter. This is because the flux increase in the X-ray spectrum is only moderate and the power-law component of the spectrum is quite comparable to the blackbody component of the spectra on most days. Comparison of the thermal \& non-thermal fluxes and the calculated MDP in the two energy bands are shown in Figure~\ref{fig8}. 

\clearpage
\subsection{H 1743-322}
\begin{figure}
\centering
\includegraphics[scale=0.37,angle=-90]{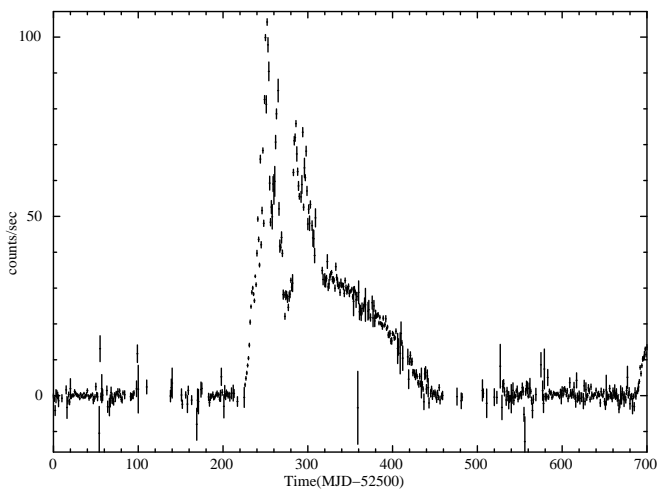}
\caption{ One day averaged long term ASM light curve of H $1743-322$ showing the $ 2003 $ outburst of H $1743-322$}
\label {fig9}
\end{figure}

\begin{figure}
\centering
\includegraphics[width=5.5cm,angle=-90]{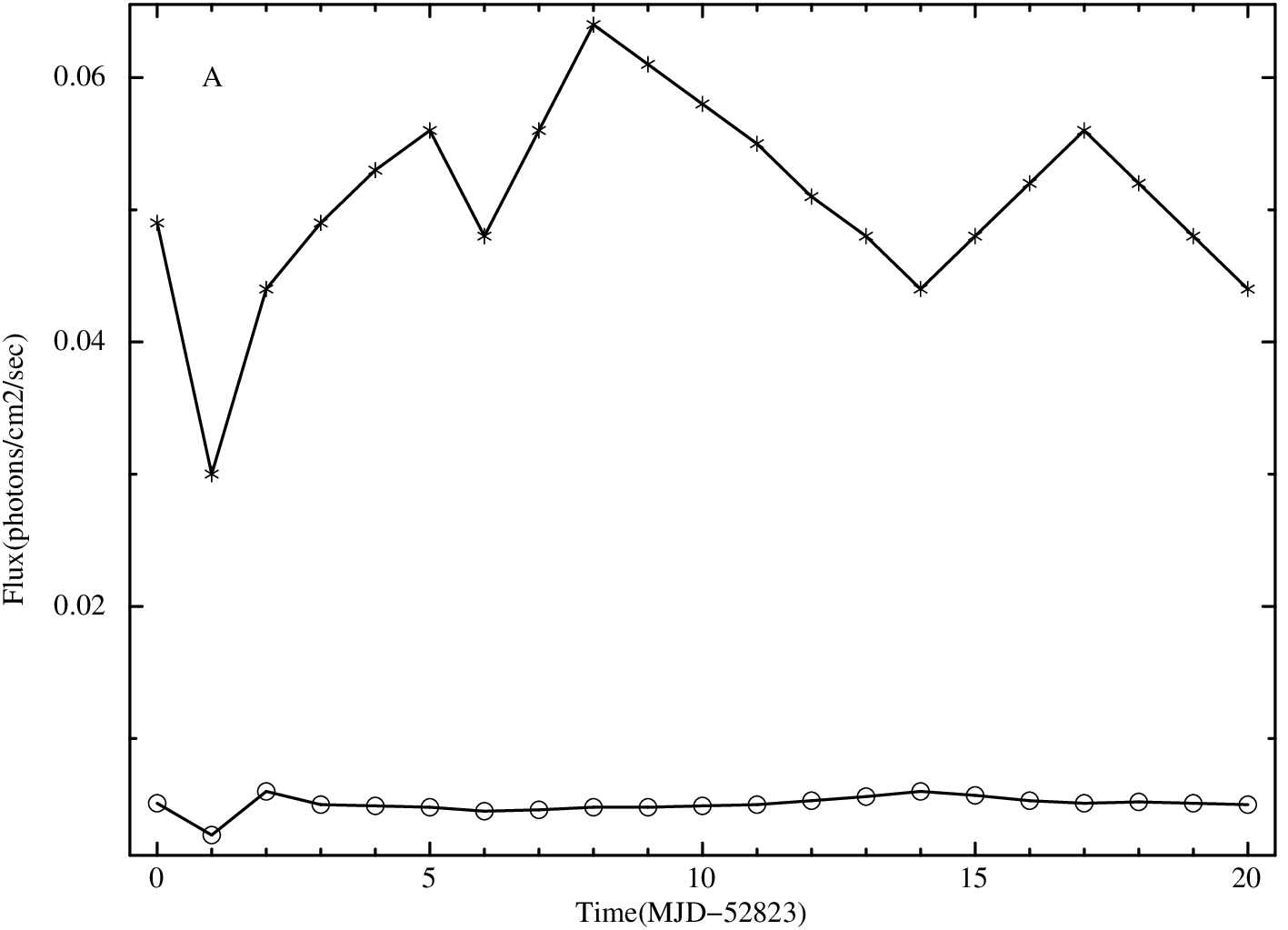}
\includegraphics[width=5.5cm,angle=-90]{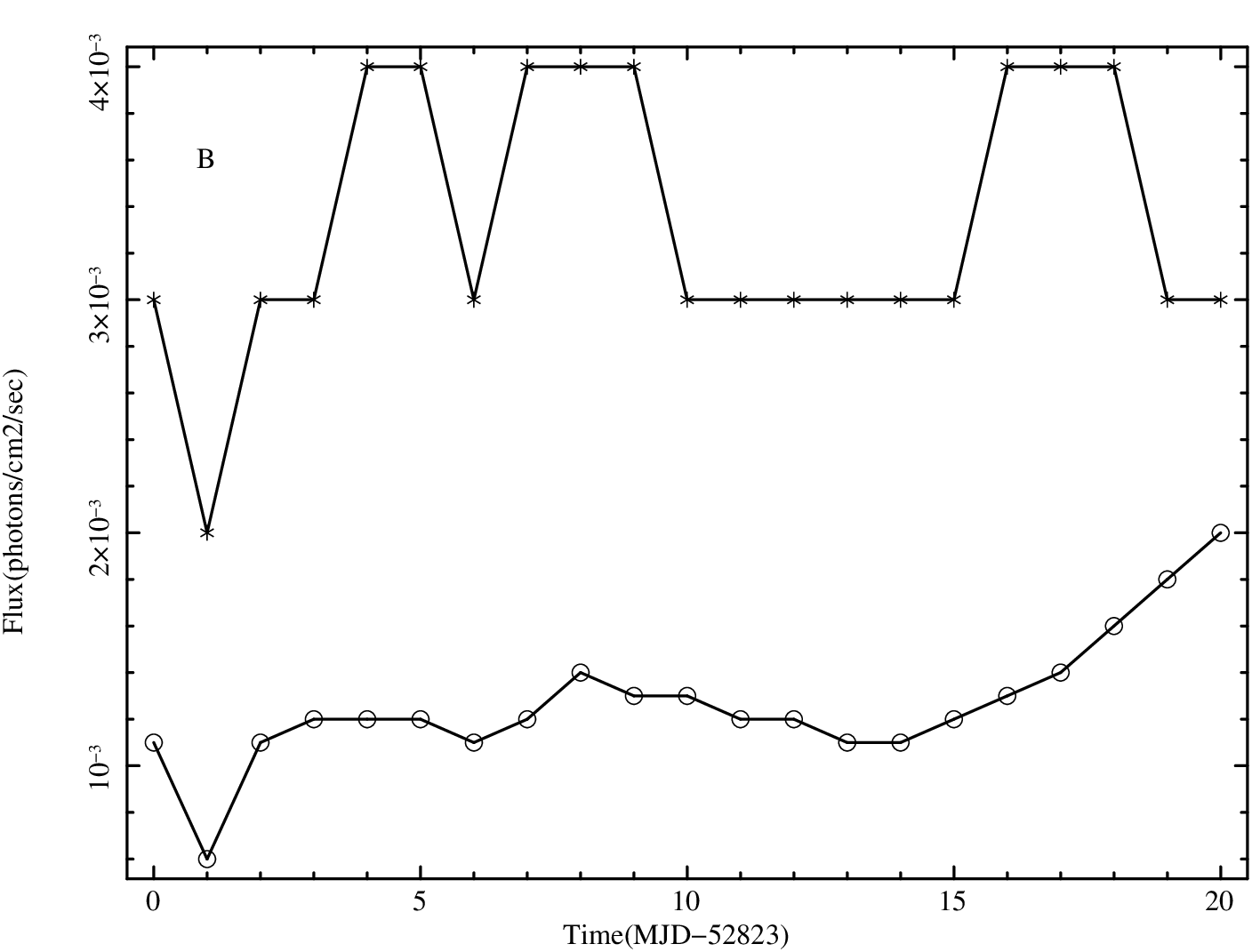}
\includegraphics[width=5.5cm,angle=-90]{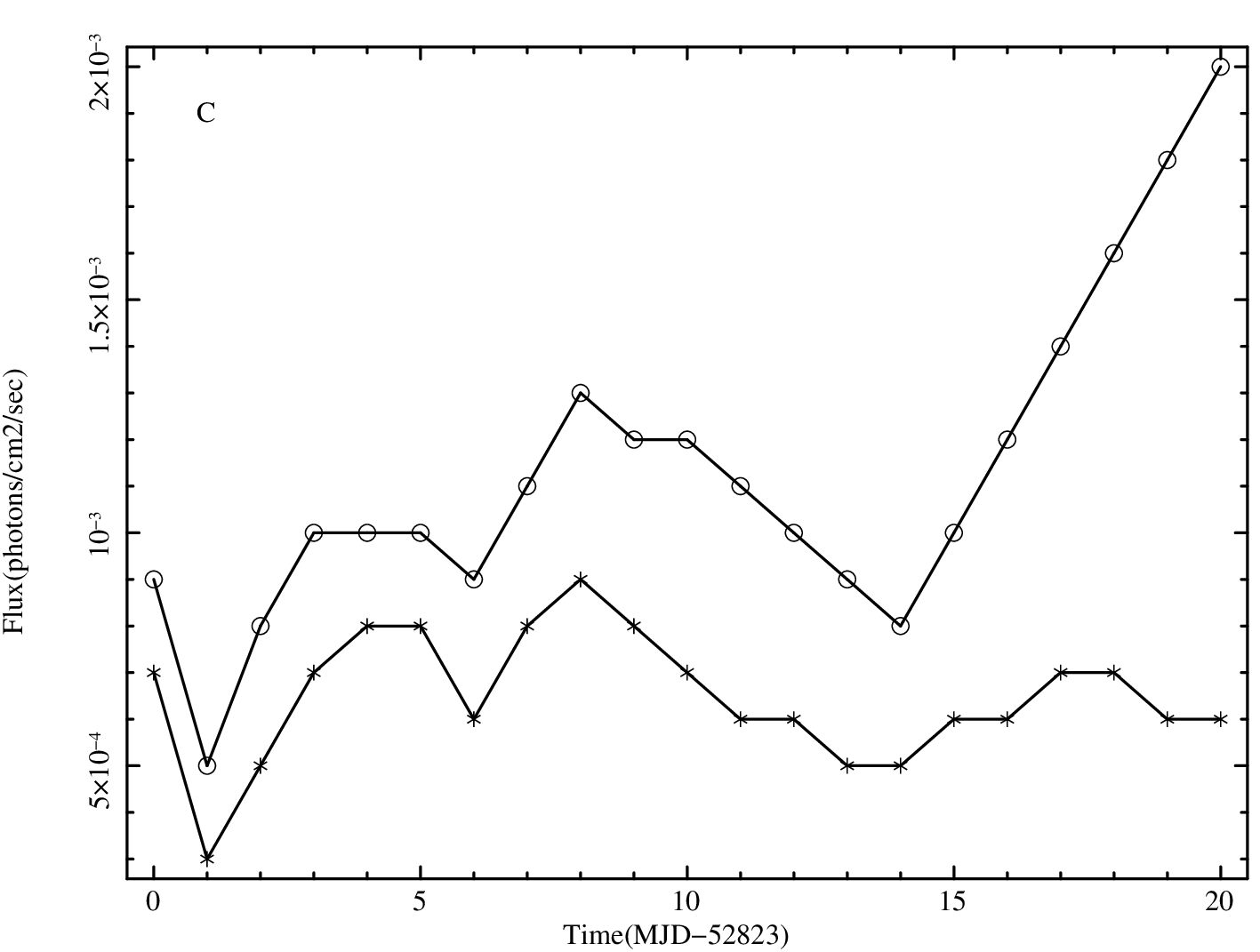}
\includegraphics[width=5.5cm,angle=-90]{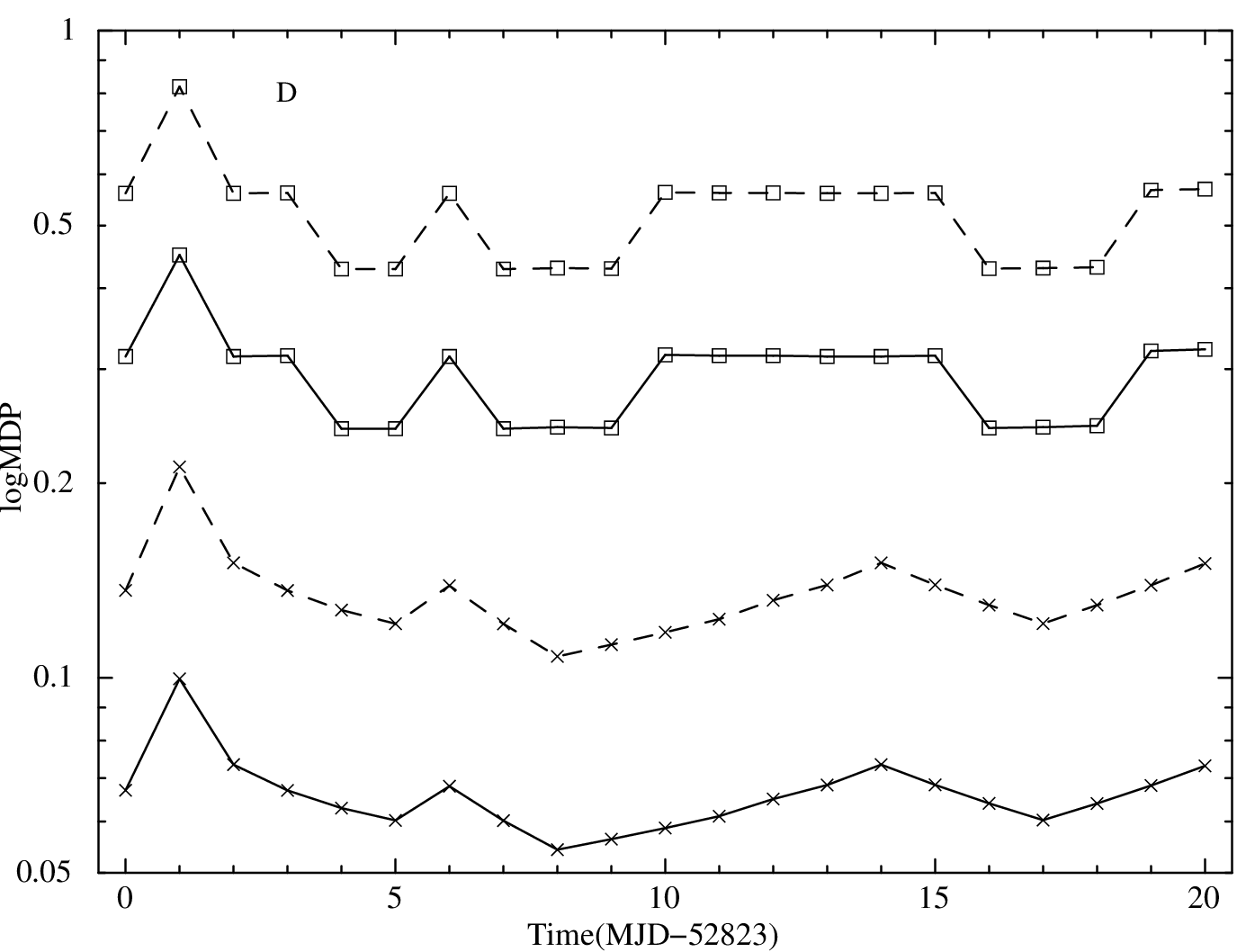}
\caption{ Same as Figure 3 for the source H1743-322 for the outburst in the year $2003$}
\label{fig10}
\end{figure}

H 1743-322 is a bright X-ray transient. It was first discovered with Ariel V all sky monitor in $1977$ \citep {kaluzienski1977} and its precise position was provided by HEAO 1. Despite lack of dynamical confirmation of the mass of the compact object it is believed to be a black hole due to its spectral and temporal characteristics \citep {corbel2006, kalemci2006}. The source also shows relativistic jet emission. The source entered into one of its outbursts in $2003$ during which it showed state transitions. The outburst was discovered by \citet {revnivtsev2003} on March $21$ using \emph{INTEGRAL} and rediscovered a few days later by \citep {markwardt2003}, and \emph{RXTE} observations of the source was carried out during the outburst. Figure~\ref{fig9} shows the one day averaged ASM light curve of this source during its outburst.\newline\\The data was analysed from MJD $ 52\,823 $ to $ 52\,843 $ to calculate the MDP. Figure~\ref{fig10} shows the comparison of the two fluxes and the MDP in the three energy ranges. 

\subsection{XTE J1817-330}
\begin{figure}
\centering
\includegraphics[scale=0.40,angle=-90]{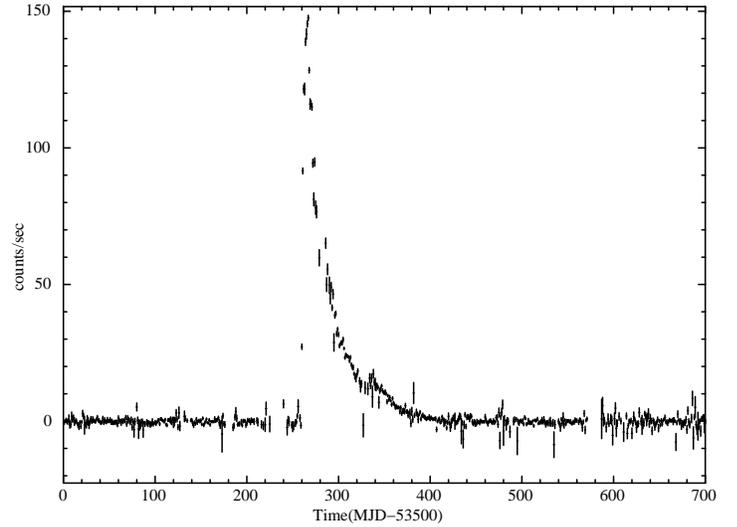}
\caption{ One day averaged long term ASM light curve of XTE J$1817-330$ showing the $2006 $ outburst of XTE J$1817-330$}
\label {fig11}
\end{figure}

\begin{figure}
\centering
\includegraphics[width=5.5cm,angle=-90]{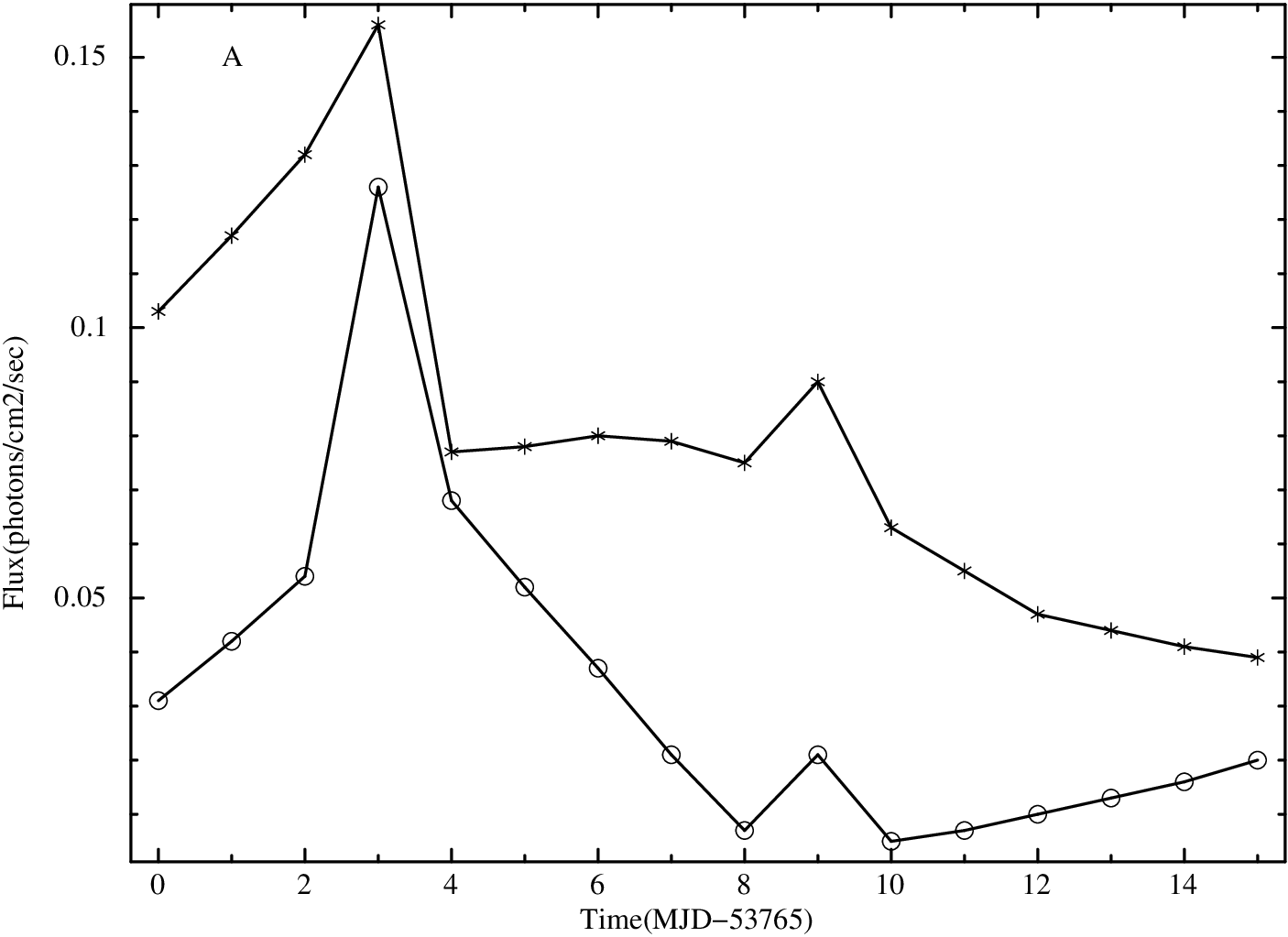}
\includegraphics[width=5.5cm,angle=-90]{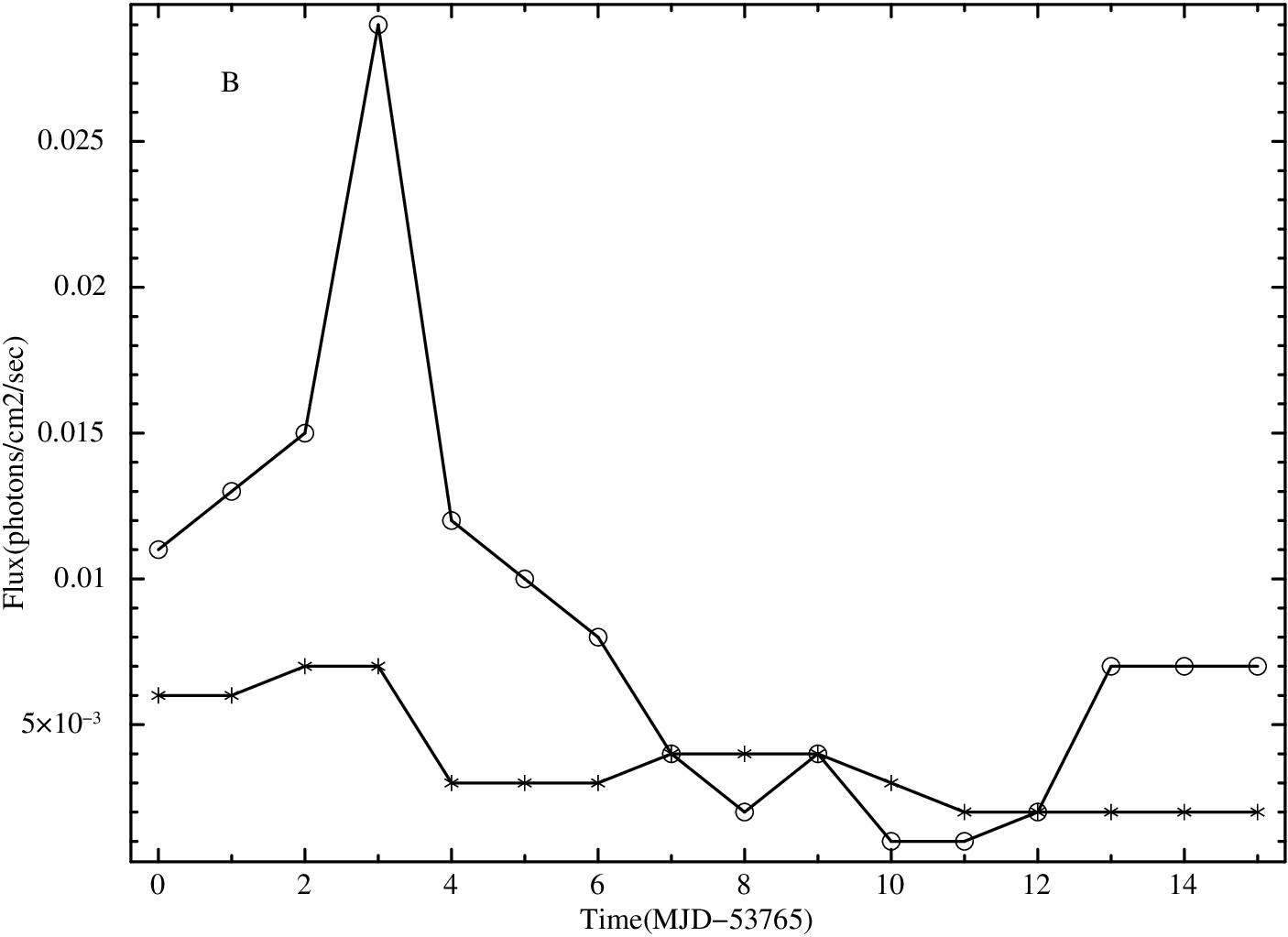}
\includegraphics[width=5.5cm,angle=-90]{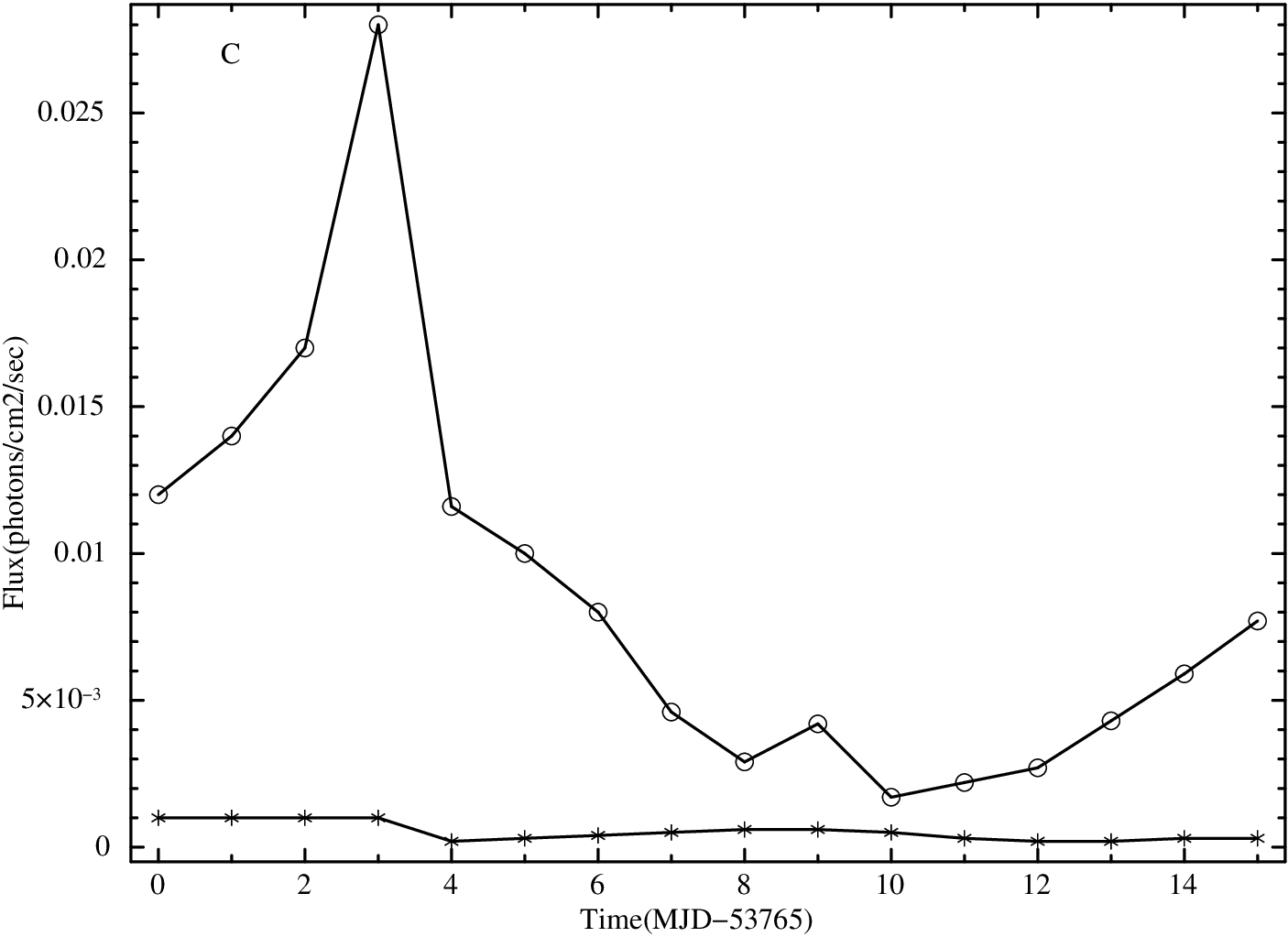}
\includegraphics[width=5.7cm,angle=-90]{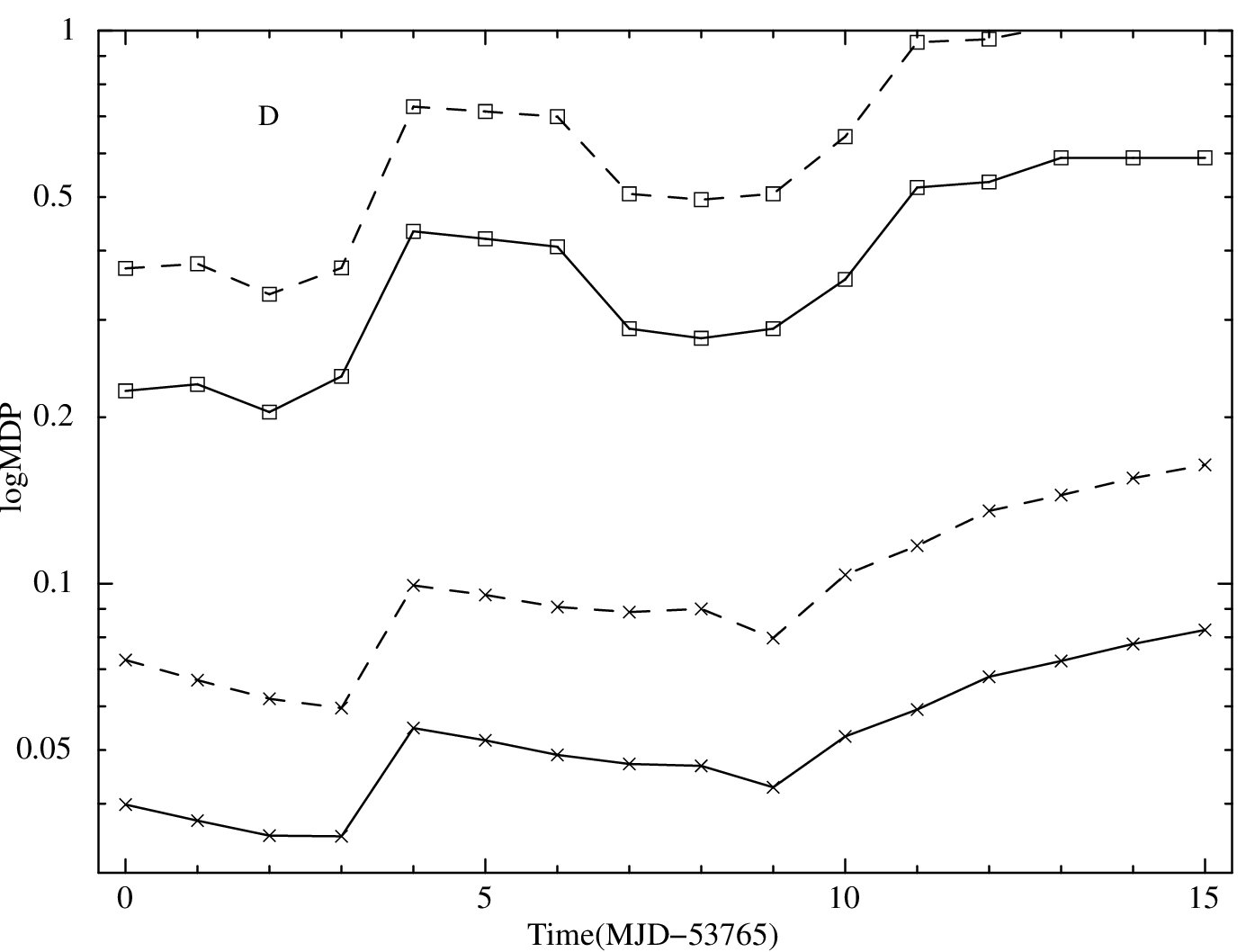}
\caption{Same as Figure 3 for the source XTE J1817-330 for the outburst in the year $2006$}
\label {fig12}
\end{figure}

XTE J1817-330 was discovered as a new bright X-ray transient with the RXTE All-Sky Monitor (ASM) on $2006$ January $26$ \citep {remillard2006} with a flux of $ 0.93 (\pm 0.03) $ Crab ($ 2-12 $ \rmfamily{KeV}) and a very soft spectrum thus making it a strong black hole candidate \citep {white1984}. Pointed observations with \emph{RXTE} confirmed the soft spectrum and also hinted low absorption along the line of sight \citep {miller2006}. Five days after the discovery of the source, its radio counterpart was identified \citep {rupen2006}, followed by detection of counterparts at NIR, optical, and ultraviolet wavelengths \citep {davanzo2006,torres2006a, torres2006b}. The $2006$ outburst lasted for approximately $ 160 $ days where the source reached a maximum of $ \sim 1.9 $ Crab on $ 28 $ January $ 2006 $, following an exponential decline of the X-ray flux. Figure~\ref{fig11} shows the one day averaged ASM daily light curve of the source during its outburst.\newline\\To calculate The MDP the data was analysed between MJD $ 53\,765 $ to $ 53\,780 $ when \emph{RXTE} pointed observations were available and the source was in  blackbody dominated days. The $ \mathrm{MDP}_{net} $ integrated over all the days was found to be similar to the other sources in $ 6-9 $ \rmfamily{KeV} \& $ 9-11 $ \rmfamily{KeV} but much worse in the $ 11-15 $ \rmfamily{KeV} energy band since the power-law flux was significantly higher and totally dominant over the blackbody flux over this energy band. Figure~\ref{fig12} shows the comparison of the two fluxes and the MDP in the specified energy ranges on each day during this time.\newline\\ Calculated values of $ \mathrm{MDP}_{net} $ for all the sources during the respective stretches are given in Table $ 2 $ as follows -\newline\\
\begin{table*} 
\centering
\begin{minipage}{140mm}
\caption{Net Minimum detectable Polarisation calculated for all the sources during the specified stretches. The values are given in percentages}
\begin{tabular}{c c c c c}
\hline \hline
Source name & Net MDP 6-9 \rmfamily{KeV} & Net MDP 9-11 \rmfamily{KeV} & Net MDP 11-15 \rmfamily{KeV} & Net MDP 11-15 \rmfamily{KeV} \\
 MJD    & $5 \sigma $ & $5 \sigma$ & $ 5 \sigma $ & $ 3 \sigma $\\
  range  & Li \quad Be & Li \quad Be & Li \quad Be & Li \quad Be \\
\hline
GRO J1655-40(53442 - 53449) & 0.97 \quad 1.69 & 2.96 \quad 4.71 & 8.37 \quad 11.63 & 5.37 \quad 7.47\\
GRO J1655-40(53511 - 53525) & 0.52 \quad 0.88 & 1.77 \quad 2.69 & 4.43 \quad 5.94 & 2.85 \quad 3.81\\
GX339-4(52472 - 52518) & 0.36 \quad 0.62 & 2.92 \quad 4.67 & 18.10 \quad 25.10 & 11.62 \quad 16.11 \\ 
Cyg X-1(50225 - 50242, 50250 - 50252) & 1.57 \quad 2.70 & 15.16 \quad 23.84 & - \quad - & - \quad -\\ 
H1743-322(52823 - 52843) & 1.42 \quad 2.89 & 6.25 \quad 11.10 & 35.94 \quad 53.63 & 23.08 \quad 34.43\\
XTE J1817-330(53765 - 53780) & 1.21 \quad 2.26 & 08.28 \quad 14.14 & 55.76 \quad - & 35.79 \quad 51.64\\ 
\hline
\end{tabular}
\end{minipage}
\end{table*}

\section{Determination of the error in the angle of polarisation measurement}
Since polarisation from accreting black holes in their thermally dominated state is also expected to have a rotation in the net angle of polarisation in the inner parts of the accretion disk as discussed previously, determination of the error or uncertainty in the angle of polarisation measurement is also necessary to specify how well one would be able to constrain the angle of polarisation measurement for a specified $ \mathrm{n}\sigma $ statistical detection. \newline To determine the error in the angle of polarisation measurement we generated a modulated signal riding on a background superimposed with random errors as would be expected to obtain from the actual measurement of a polarised signal with the Thomson polarimeter. Adjusting the amplitude and background to fix the signal at specified $ \mathrm{n}\sigma $ levels of detection we determined the broadening in the phase (angle of polarisation) measurement \rmfamily{i.e.} the $ 1\sigma $ uncertainty in the polarisation angle measurement at each $ \mathrm{n}\sigma$ level. Similar results have been found by  M. Weisskopf et al. (2010). Figure $14 $ shows the $ 1\sigma $ deviation in phase versus the number of sigma $ (\sigma) $ detection.  \newline\\
\begin{figure}
\centering
\includegraphics[scale=0.38,angle=-90]{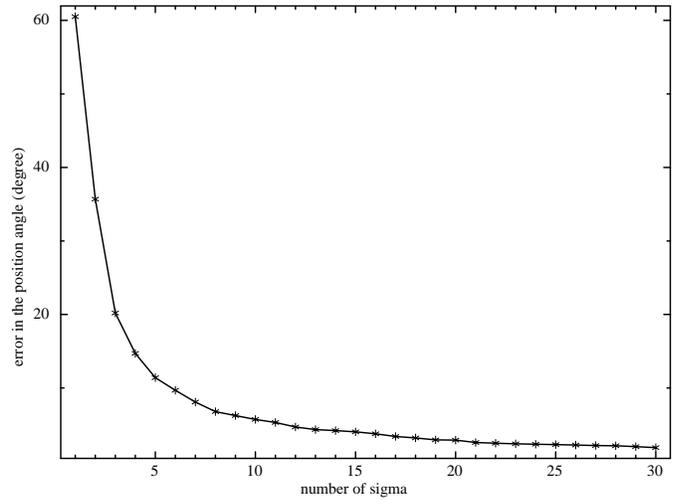}
\caption{The 1$\sigma $ deviation in position angle versus the number of sigma statistical detection }
\label {fig13}
\end{figure}
\section{Discussion \& Conclusions}
We have shown that provided some of the sources go into similar outburst/state transition during the lifetime of the mission, detection of polarisation in the disc emission would be possible with the Thomson polarimeter. This can be done by performing spectro polarimetric observations in different energy bands to map the energy dependent polarisation for these sources in their disc emission dominated state.\\X-ray emission from the disk in accreting black holes are polarised due to electron scattering from the plane parallel atmosphere of the disk with the polarisation vector lying in the plane of the disk. However, relativistic effects like beaming and gravitational lensing gives a non-trivial rotation to the integrated net polarisation vector resulting in energy dependent rotation in the plane of polarisation. Recent Monte Carlo Ray Tracing simulations \citep {schnittman2009} predict a decrease in the degree of polarisation along with its rotation in the inner parts of the accretion disk for direct radiation, and a net rotation of ninety degrees in the plane of polarisation in the same region considering returning radiation. They have also simulated the results that would be expected with the polarisation measurement taken by the \emph{GEMS} mission polarimeter which is a photoelectron polarimeter working in the $ 2 - 10 $ \rmfamily{KeV} range. Here we produce the results that we would expect to get with the Thomson polarimeter. The results could also be complimented with the \emph{GEMS} enabling extensive wideband spectro polarimetric studies. The main conclusions of this paper are:
\begin{enumerate}
\item The net MDP obtained for these sources in the three energy bands of $ 6 - 9 $ \rmfamily{KeV}, $ 9 - 11 $ \rmfamily{Kev} \& $ 11 - 15 $ \rmfamily{KeV} is in most cases less than the degree of polarisation expected from accreting black holes in their thermal state from the simulations of \citep {schnittman2009}, the degree and the net rotation in the integrated polarisation vector being a strong function of the black hole spin. A comparison of the net MDP presented in Table 2 for three different energy bands with the predicted degree of polarisation (\citet {schnittman2009}, Figures 7 \& 8) shows that for sources with favourable inclination angle (\emph{i}$ > $75 degrees) spectro-polarimetric measurements will be possible with the proposed Thomson X-ray polarimeter and these measurements will also be useful to constrain the spin parameter of the black hole. In some cases however, where the spectrum is almost entirely powerlaw dominated in the $11-15 $ \rmfamily{KeV} band, the MDP obtained is above the the degree of polarisation expected to be detected from such sources in case of a $5 \sigma $ statistical detection. A lower significance detection however, may be possible in such cases as a $ 3 \sigma $ detection gives a lower achievable value of MDP. The calculated error in the angle of polarisation obtained for a $ 3 \sigma $ statistical detection onwards is also within reasonable limits. Hence spectro polarimetric observation of such sources for a few weeks during their thermally dominated spectral state would possibly enable us to map the transition radius from horizontal to vertical polarisation. This could provide us with some estimates on the parameters of the black holes like their spin and/or the emissivity profile.
\item Observation of polarisation in the $ 6 - 9 $ \rmfamily{KeV} energy range necessitates the use of lithium as the scattering element which would set the dynamic range of the experiment to $ 5 - 30 $ \rmfamily{KeV}. Use of beryllium however limits the lower energy energy threshold to  $ 8 $ \rmfamily{keV} thus making observation in the $ 6 - 9 $ \rmfamily{keV} band having the lowest MDP impossible. This study thus strongly supports the use of lithium as the scattering element for the Thomson polarimeter.
   \end{enumerate}
\section*{Acknowledgments}
This research has made use of data obtained through the High Energy Astrophysics Science Archive Research Center On line Service, provided by NASA/Goddard Space Flight Center.

\label{lastpage}

\end{document}